\documentclass[aip,rsi, amsmath,amssymb,reprint]{revtex4-1}
\usepackage{siunitx} 
\usepackage{graphicx}
\usepackage[utf8]{inputenc}
\usepackage[english]{babel}
\usepackage{blindtext}
\usepackage{enumitem}
\usepackage{sidecap}
\usepackage{hyperref}
\draft % marks overfull lines with a black rule on the right

\begin{document}
%\preprint{}

\title{An ultralow-noise superconducting radio-frequency ion trap for frequency metrology with highly charged ions} %Title of paper



\author{J. Stark}
\email[Author to whom correspondence should be addressed: ]{julian.stark@mpi-hd.mpg.de}
\affiliation{Max-Planck-Institut f\"ur Kernphysik, Saupfercheckweg 1, 69117 Heidelberg, Germany}
\affiliation{Heidelberg Graduate School for Physics, Ruprecht-Karls-Universität Heidelberg, Im Neuenheimer Feld 226, 69120 Heidelberg, Germany}

\author{C. Warnecke}
\affiliation{Max-Planck-Institut f\"ur Kernphysik, Saupfercheckweg 1, 69117 Heidelberg, Germany}
\affiliation{Heidelberg Graduate School for Physics, Ruprecht-Karls-Universität Heidelberg, Im Neuenheimer Feld 226, 69120 Heidelberg, Germany}

\author{S. Bogen}
\affiliation{Max-Planck-Institut f\"ur Kernphysik, Saupfercheckweg 1, 69117 Heidelberg, Germany}

\author{S. Chen}
\affiliation{Max-Planck-Institut f\"ur Kernphysik, Saupfercheckweg 1, 69117 Heidelberg, Germany}
\affiliation{State Key Laboratory of Magnetic Resonance and Atomic and Molecular Physics, Wuhan Institute of Physics and Mathematics, Innovation Academy for Precision Measurement Science and Technology, Chinese Academy of Sciences, Wuhan 430071, China}

\author{E. A. Dijck}
\affiliation{Max-Planck-Institut f\"ur Kernphysik, Saupfercheckweg 1, 69117 Heidelberg, Germany}

\author{S. Kühn}
\affiliation{Max-Planck-Institut f\"ur Kernphysik, Saupfercheckweg 1, 69117 Heidelberg, Germany}
\affiliation{Heidelberg Graduate School for Physics, Ruprecht-Karls-Universität Heidelberg, Im Neuenheimer Feld 226, 69120 Heidelberg, Germany}

\author{M. K. Rosner}
\affiliation{Max-Planck-Institut f\"ur Kernphysik, Saupfercheckweg 1, 69117 Heidelberg, Germany}
\affiliation{Heidelberg Graduate School for Physics, Ruprecht-Karls-Universität Heidelberg, Im Neuenheimer Feld 226, 69120 Heidelberg, Germany}

\author{A. Graf}
\affiliation{Max-Planck-Institut f\"ur Kernphysik, Saupfercheckweg 1, 69117 Heidelberg, Germany}

\author{J. Nauta}
\affiliation{Max-Planck-Institut f\"ur Kernphysik, Saupfercheckweg 1, 69117 Heidelberg, Germany}
\affiliation{Heidelberg Graduate School for Physics, Ruprecht-Karls-Universität Heidelberg, Im Neuenheimer Feld 226, 69120 Heidelberg, Germany}

\author{J.-H. Oelmann}
\affiliation{Max-Planck-Institut f\"ur Kernphysik, Saupfercheckweg 1, 69117 Heidelberg, Germany}
\affiliation{Heidelberg Graduate School for Physics, Ruprecht-Karls-Universität Heidelberg, Im Neuenheimer Feld 226, 69120 Heidelberg, Germany}

\author{L. Schmöger}
\affiliation{Max-Planck-Institut f\"ur Kernphysik, Saupfercheckweg 1, 69117 Heidelberg, Germany}
\affiliation{Physikalisch-Technische Bundesanstalt, Bundesallee 100, 38116 Braunschweig, Germany}

\author{M. Schwarz}
\affiliation{Max-Planck-Institut f\"ur Kernphysik, Saupfercheckweg 1, 69117 Heidelberg, Germany}
\affiliation{Physikalisch-Technische Bundesanstalt, Bundesallee 100, 38116 Braunschweig, Germany}

\author{D. Liebert}
\affiliation{Max-Planck-Institut f\"ur Kernphysik, Saupfercheckweg 1, 69117 Heidelberg, Germany}

\author{L. J. Spieß}
\affiliation{Max-Planck-Institut f\"ur Kernphysik, Saupfercheckweg 1, 69117 Heidelberg, Germany}
\affiliation{Physikalisch-Technische Bundesanstalt, Bundesallee 100, 38116 Braunschweig, Germany}

\author{S. A. King}
\affiliation{Physikalisch-Technische Bundesanstalt, Bundesallee 100, 38116 Braunschweig, Germany}

\author{T. Leopold}
\affiliation{Physikalisch-Technische Bundesanstalt, Bundesallee 100, 38116 Braunschweig, Germany}

\author{P. Micke}
\affiliation{Max-Planck-Institut f\"ur Kernphysik, Saupfercheckweg 1, 69117 Heidelberg, Germany}
\affiliation{Physikalisch-Technische Bundesanstalt, Bundesallee 100, 38116 Braunschweig, Germany}

\author{P. O. Schmidt}
\affiliation{Physikalisch-Technische Bundesanstalt, Bundesallee 100, 38116 Braunschweig, Germany}
\affiliation{Institut für Quantenoptik, Leibniz Universität Hannover, Welfengarten 1, 30167 Hannover, Germany}

\author{T. Pfeifer}
\affiliation{Max-Planck-Institut f\"ur Kernphysik, Saupfercheckweg 1, 69117 Heidelberg, Germany}

\author{J. R. Crespo L\'{o}pez-Urrutia}
\affiliation{Max-Planck-Institut f\"ur Kernphysik, Saupfercheckweg 1, 69117 Heidelberg, Germany}

\date{\today}

\begin{abstract}
We present a novel ultrastable superconducting radio-frequency (RF) ion trap realized as a combination of an RF cavity and a linear Paul trap.
Its RF quadrupole mode at $\SI{34.52}{\mega\hertz}$ reaches a quality factor of $Q\approx2.3\times 10^5$ at a temperature of $\SI{4.1}{\kelvin}$ and is used to radially confine ions in an ultralow-noise pseudopotential.
This concept is expected to strongly suppress motional heating rates and related frequency shifts which limit the ultimate accuracy achieved in advanced ion traps for frequency metrology.
Running with its low-vibration cryogenic cooling system, electron beam ion trap and
deceleration beamline supplying highly charged ions (HCI), the superconducting trap offers ideal conditions for optical frequency metrology with ionic species.
%The superconducting ion trap is cooled to cryogenic temperatures by a low-vibration closed-cycle system and offers ideal conditions for optical frequency metrology, in particular with highly charged ions (HCI).
%They are provided by a dedicated link to an electron beam ion trap through a deceleration beamline.
We report its proof-of-principle operation as a quadrupole mass filter with HCI, and trapping of Doppler-cooled ${}^9\text{Be}^+$ Coulomb crystals.

\end{abstract}
\pacs{}% insert suggested PACS numbers in braces on next line

\maketitle %\maketitle must follow title, authors, abstract and \pacs

%%%%%%%%%%%%%%%%%%%%%%%%%%%%%%%%%%%%%%%%%%%%%%%%%%%%%%%%%%%%%%%%%%%%%%%%%%%%
%%%%%%%%%%%%%%%%%%%%%%%%%%%%%%%%%%%%%%%%%%%%%%%%%%%%%%%%%%%%%%%%%%%%%%%%%%%%
\section{Introduction}\label{sec:I}
Over the past decades, Paul traps have proven themselves as indispensable instruments in physics and chemistry, as well as wide-spread analytical applications.
Their confinement of ions inside a zero-field environment with long storage times makes them especially suited for quantum computing and optical frequency metrology:\cite{wineland_experimental_1998, poli_optical_2013, ludlow_optical_2015} High trapping frequencies allow for recoil-free absorption of photons, enabling quantum computation\cite{leibfried_quantum_2003, blatt_entangled_2008} and quantum logic spectroscopy\cite{schmidt_spectroscopy_2005} (QLS) by coupling electronic and motional degrees of freedom of the ions.
Crucially, this has paved the way for many fundamental physics studies with atomic systems (for a review see, e.g., Ref.\ \onlinecite{Safronova_Search_NP_2018}), such as searches for a possible temporal variation of fundamental constants\cite{rosenband_frequency_2008, huntemann_improved_2014} or local Lorentz invariance,\cite{pruttivarasin_michelsonmorley_2015,megidish_improved_2019, sanner_optical_2019} that have been made possible by the ultimate accuracy and low systematic uncertainties of Paul trap experiments.\cite{huntemann_single-ion_2016,keller_controlling_2019}

For such fundamental studies, highly charged ions (HCI) are very interesting candidates (see, e.g., Ref. \onlinecite{kozlov_highly_2018}).
Due to the steep scaling of their binding energies with charge state, fine-structure and hyperfine-structure transitions can be shifted to the optical range and become reachable for high-precision laser spectroscopy.\cite{berengut_highly_2012}
In addition, transitions between energetically close electronic configurations at level crossings are found to be in the optical range. \cite{berengut_enhanced_2010, bekker_detection_2019}
HCI have been proposed to test Standard Model extensions, as some HCI feature electronic transitions with enhanced sensitivity to a possible variation of fundamental constants, \cite{schiller_hydrogenlike_2007,berengut_electron-hole_2011, berengut_optical_2012,Derevianko_HCI_exceptional_2012,dzuba_ion_2012, kozlov_highly_2018} or to probe new spin-independent long-range interactions using isotope shift measurements with the generalized King plot method.\cite{yerokhin_nonlinear_2020, berengut_generalized}
Resulting from the compact size of their electronic orbitals, HCI feature reduced atomic polarizabilities, small electric quadrupole moments, and suppressed field-induced shifts.\cite{kozlov_highly_2018}
This also renders them promising candidates \cite{schiller_hydrogenlike_2007,dzuba_ion_2012, Derevianko_HCI_exceptional_2012, yudin_magnetic-dipole_2014,kozlov_highly_2018} for next-generation frequency standards with suggested relative systematic uncertainties below $10^{-19}$.
In addition, HCI offer forbidden transitions in the ultraviolet (UV), vacuum ultraviolet (VUV) and soft x-ray regions,\cite{Schuessler2020} which allow the development of ion-based frequency standards with improved stability.\cite{kozlov_highly_2018}

What used to be called `HCI precision experiments' were for several decades carried out with electron beam ion traps (EBIT), ion-storage rings, and electron-cyclotron ion sources (e.g.\ Refs.\ \onlinecite{Beiersdorfer1998, Brandau2003, draganic_high_2003, soria_orts_zeeman_2007, mackel_laser_2011, PhysRevLett.109.043005, PhysRevA.97.032517}).
Due to the high motional ion temperatures ($T>\SI{e5}{\kelvin}$) in those devices, the achieved relative spectral resolution merely reached the parts-per-million level.
A few years ago, sympathetic cooling of HCI transferred from an EBIT into a cryogenic Paul trap containing a laser-cooled Coulomb crystal of ${}^9\text{Be}^+$ ions brought down the accessible temperatures of trapped HCI from the megakelvin into the millikelvin range.\cite{schmoger_coulomb_2015}
Later, a pioneering experiment in a Penning trap performed spectroscopy on the forbidden optical fine-structure transition in ${}^{40}\text{Ar}^{13+}$ at $\SI{441}{\nano\metre}$ at higher ion temperatures ($T\approx 1\SI{}{\kelvin}$) by laser-induced excitation and subsequent detection through a measurement of the electronic ground-state spin orientation,\cite{egl_application_2019} reaching a relative uncertainty of $\Delta\nu/\nu\simeq 9.4\times 10^{-9}$.
The potential of HCI for optical frequency metrology was finally unleashed with the recent ground-state cooling of the axial modes of motion ($T<\SI{50}{\micro\kelvin}$) of a two-ion crystal consisting of one HCI and one $^9\text{Be}^+$ ion and the subsequent application of QLS in a Paul trap to the aforementioned forbidden transition in ${}^{40}\text{Ar}^{13+}$, reporting a statistical uncertainty of $\Delta\nu/\nu\simeq 10^{-15}$.\cite{micke_coherent_2020}

One major effect that systematically limits the achievable accuracy in Paul trap experiments is the time-dilation shift caused by residual ion motion, which represents a key problem for frequency standards based on singly charged ions.\cite{brewer_quantum-logic_2019}
To overcome this limitation requires a strong reduction of trap-induced heating rates, preserving small occupation numbers of the quantum harmonic oscillator throughout the interrogation time.
%Alternatively, active sympathetic cooling during clock interrogation can be employed.\cite{brewer_quantum-logic_2019}

In this paper, we present a new cryogenic Paul trap experiment providing ultrastable trapping conditions which promises an exceptionally high suppression of motional heating rates.
Its centerpiece is a novel radio-frequency (RF) ion trap realized by integrating a linear Paul trap into a quasi-monolithic superconducting RF cavity.
The resonant quadrupole mode (QM) of its electric field generates an ultrastable pseudopotential that radially confines the ions.
Originally developed for experiments with HCI, which are expected to exhibit higher heating rates than singly charged ions,\cite{brownnutt_ion-trap_2015, king_algorithmic_2021} the technique can also be applied to any other ion species.

The cryogenic and vacuum setup builds upon the Cryogenic Paul Trap Experiment (CryPTEx)\cite{schwarz_cryogenic_2012} at the Max-Planck-Institut für Kernphysik (MPIK) in Heidelberg and on the cryogenic design of Refs.\ \onlinecite{leopold_cryogenic_2019,micke_closed-cycle_2019} developed at MPIK in collaboration with the Physikalisch-Technische Bundesanstalt (PTB) in Braunschweig for the CryPTEx-PTB\cite{leopold_cryogenic_2019, micke_coherent_2020} experiment, and is consequently named CryPTEx-SC.\cite{stark_phd_2020}
The low-vibration cryogenic supply\cite{micke_closed-cycle_2019} provides mechanically ultrastable trapping conditions by decoupling external vibrations from the trap region.
An EBIT\cite{micke_heidelberg_2018} is added as a source for HCI, and a low-energy HCI transfer beamline\cite{micke_be_2020, rosner_2019} connects EBIT and Paul trap.
Since we aim for direct frequency comb spectroscopy of HCI in the extreme ultraviolet (XUV) range, a dedicated XUV frequency comb based on high-harmonic-generation inside an optical enhancement cavity has been set up and commissioned at MPIK.\cite{nauta_towards_2017, Nauta_VMI_2020, nauta2020xuv}

%%%%%%%%%%%%%%%%%%%%%%%%%%%%%%%%%%%%%%%%%%%%%%%%%%%%%%%%%%%%%%%%%%%%%%%%%%%%
%%%%%%%%%%%%%%%%%%%%%%%%%%%%%%%%%%%%%%%%%%%%%%%%%%%%%%%%%%%%%%%%%%%%%%%%%%%%
\section{Concept}\label{Sec_Concept}
%%%%%%%%%%%%%%%%%%%%%%%%%%%%%%%%%%%%%%%%%%%%%%%%%%%%%%%%%%%%%%%%%%%%%%%%%%%%
\begin{figure}[t]
\centering
\includegraphics[width=\columnwidth]{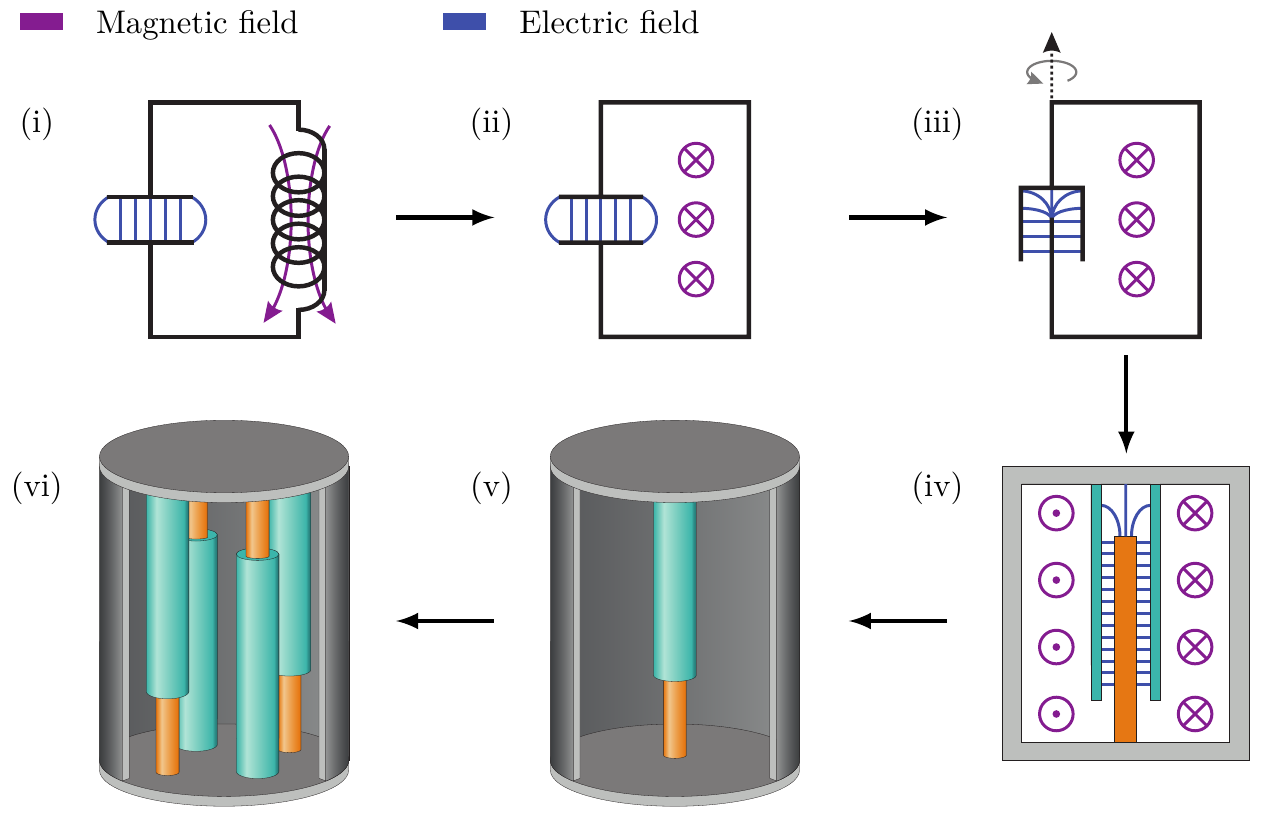}
\caption{Principle of a cavity with an electric QM.
A two-dimensional ideal \textit{LC} circuit (i-iii) rotating around the specified axis generates a three-dimensional dipole resonator (v) with coaxial cross section (iv).
Four symmetrically arranged resonator poles with alternating polarity (vi) then shape a quadrupole electrode structure inside a single resonant tank.
Reprinted from Ref.\ \onlinecite{stark_phd_2020}.}
\label{Fig_LC_Schematic_Working_Principle}
\end{figure}
%%%%%%%%%%%%%%%%%%%%%%%%%%%%%%%%%%%%%%%%%%%%%%%%%%%%%%%%%%%%%%%%%%%%%%%%%%%%

The superconducting cavity (SCC) generating the pseudopotential for ion confinement integrates a linear Paul trap as shown schematically in Fig.\ \ref{Fig_LC_Schematic_Working_Principle}.
The resonance frequency $\omega_0$ of its electric QM is identical to the trap drive frequency, $\Omega \equiv \omega_0$.
The ions are confined in a superposition of the thereby generated 2D pseudopotential and an electrostatic potential along the third direction, which is configured by additional electrodes integrated in the quadrupole electrodes (not shown in Fig.\ \ref{Fig_LC_Schematic_Working_Principle}).
Each of the quadrupole electrodes consists of an outer shell electrode with a cylindrical bore containing a coaxial inner conductor separated by a narrow gap.
These two elements are opposite RF poles of the cavity, and their small separation increases the lumped capacitance of the cavity and thus lowers its resonance frequency.
Their relative RF phase is fully defined by geometry, which, given a manufacturing tolerance conservatively estimated to $\leq \SI{100}{\um}$, suppresses phase differences between them to the level of $\Delta \Phi \leq \SI{7.2e-5}{\radian}$.
In this way, a typical source of excess micromotion, which is otherwise difficult to compensate with wired trap configurations, is strongly reduced.\cite{berkeland_minimization_1998}

Quasi-monolithic resonators reach very high values of the quality factor $Q$, commonly defined as the ratio of the stored electromagnetic energy $W$ to the dissipated power $P_\text{d}$ per RF cycle, $Q = \omega_0W/P_\text{d}$.
In our case, the SCC strongly reduces resistive losses and increases $Q$.
As this parameter also sets the time scale $\tau = 2Q/ \omega_0$ for the decay of the stored electromagnetic energy, amplitude and phase fluctuations are averaged over many cycles in our cavity, which generates very stable values for the RF-voltage amplitude and the associated time-averaged pseudopotential.
%In addition, the high quality factor of the resonator enables high electric field amplitudes for a given input power.
%This allows for comparatively large geometric scaling factors of the trap, compared to other designs, while maintaining tight ion confinement.

More importantly, $Q$ determines the bandwidth of the RF cavity excitation spectrum, $\Delta \omega= \omega_0/Q$, and filters out noise from the external RF drive for all frequencies separated by more than a few linewidths $\Delta \omega$ from its resonance.
This should result in spectrally narrow secular frequencies and strongly reduced motional heating rates of trapped ions.
In particular, the motional mode frequencies $\omega_i \ll \omega_0$ are well separated from the QM, $|\omega_0 -\omega_i|\gg \Delta \omega $, and residual RF noise at $\omega_0 \pm \omega_i$ or $\omega_i$, that would result in ion heating,\cite{brownnutt_ion-trap_2015} is drastically suppressed.

Other noise sources causing motional heating are also strongly cancelled.
Johnson-Nyquist noise, the largest non-anomalous heating source in room temperature setups, is greatly reduced in the SCC operating at $\SI{4}{\kelvin}$.
The electrostatic trap voltages are fed through low-pass filters that are held at a temperature of $\SI{4}{\kelvin}$.
Anomalous heating, depending on the distance $r_0$ between ion and electrode as typically $r_0^{-2}$ to $r_0^{-4}$,\cite{brownnutt_ion-trap_2015} has a microscopic reach and is strongly suppressed at the distance scale $r_0=\SI{1.75}{\milli\metre}$ of this trap.

%%%%%%%%%%%%%%%%%%%%%%%%%%%%%%%%%%%%%%%%%%%%%%%%%%%%%%%%%%%%%%%%%%%%%%%%%%%
\begin{figure*}[t]
\centering
\includegraphics[width=\textwidth]{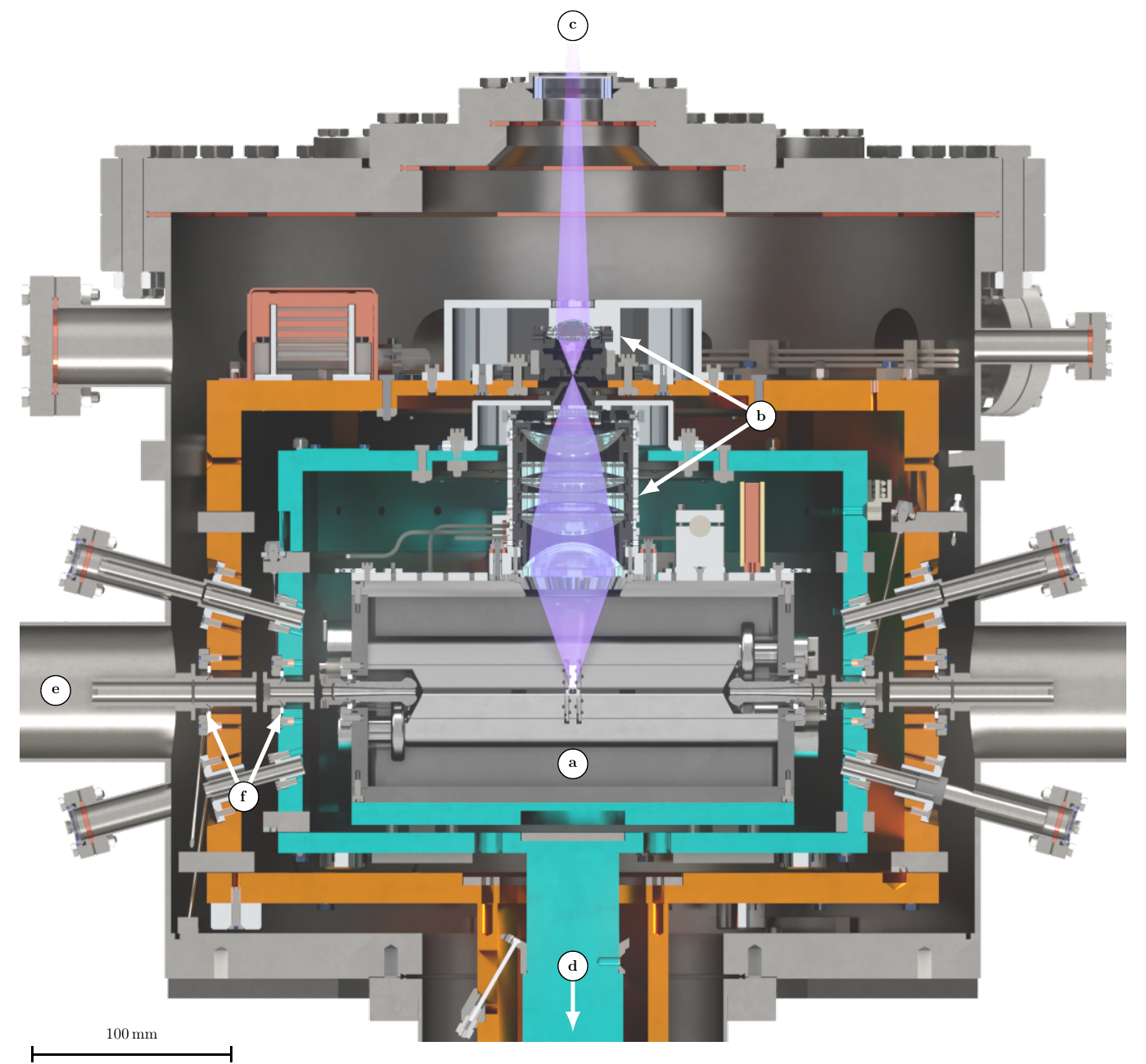}
\caption{Section through the cryogenic trap setup: first ($\SI{69}{\kelvin}$, in orange) and second ($\SI{4}{\kelvin}$, in cyan) temperature stage; (\textbf{a}) superconducting cavity, (\textbf{b}) imaging system, (\textbf{c}) detection system, (\textbf{d}) connection to cryogenic supply, (\textbf{e}) port to HCI transfer beamline, (\textbf{f}) electrostatic lenses.
Detailed views of the cavity are shown in Fig.\ \ref{Fig_CAD_model_SRF_cavity} and Fig.\ \ref{Fig_CAD_model_SRF_cavity_radial_section}.
Adapted from Ref.\ \onlinecite{stark_phd_2020}.}
\label{Fig_CAD_model_Cold_stage}
\end{figure*}
%%%%%%%%%%%%%%%%%%%%%%%%%%%%%%%%%%%%%%%%%%%%%%%%%%%%%%%%%%%%%%%%%%%%%%%%%%%%

%%%%%%%%%%%%%%%%%%%%%%%%%%%%%%%%%%%%%%%%%%%%%%%%%%%%%%%%%%%%%%%%%%%%%%%%%%%%
%%%%%%%%%%%%%%%%%%%%%%%%%%%%%%%%%%%%%%%%%%%%%%%%%%%%%%%%%%%%%%%%%%%%%%%%%%%%
\section{Design}\label{sec:II}

Long ion storage times on the order of ten minutes are needed for QLS and frequency metrology experiments.
Thus, experiments with HCI crucially depend on extremely high vacuum (XHV) conditions to suppress charge-exchange reactions with residual gas.
Here, the cryogenic trap environment reduces the pressure to levels below $\SI{e-14}{\milli\bar}$.\cite{schwarz_cryogenic_2012, pagano_cryogenic_2018, micke_closed-cycle_2019}
The three main design requirements for the ion-trap environment consisting of SCC and the surrounding cryogenic setup, shown in Fig.\ \ref{Fig_CAD_model_Cold_stage}, are: (R1) multiple optical access ports to the trap center for lasers, external atom or ion sources, and detection of fluorescence photons; (R2) efficient capturing and preparation of HCI inside the trap to optimize the measurement cycle; and (R3), a high mechanical stability and low differential contraction during cooldown to $\SI{4}{\kelvin}$ to avoid misalignment.

%%%%%%%%%%%%%%%%%%%%%%%%%%%%%%%%%%%%%%%%%%%%%%%%%%%%%%%%%%%%%%%%%%%%%%%%%%%%
\subsection{Cryogenic setup}\label{Sec_Specifications_Cavity_design}
We use a pulse-tube cryocooler (Sumitomo
Heavy Industries RP-082, specified with $\SI{40}{\watt}$ at $\SI{45}{\kelvin}$ and $\SI{1}{\watt}$ at $\SI{4}{\kelvin}$) connected to the cryogenic trap environment by means of a low-vibration supply\cite{micke_closed-cycle_2019} to refrigerate two nested thermal stages inside the vacuum chamber, where the outer stage shields the inner one from room temperature thermal radiation.
Both are made of $99.99\%$ oxygen-free high thermal conductivity (OFHC) copper.
For avoiding misalignment of the first stage with respect to the vacuum chamber and the second stage with respect to the first stage during cooldown (R3), the setup follows a symmetric design, with sets of equally long counteracting stainless-steel spokes holding the thermal stages.
In this way, thermal contraction forces stay in balance and keep the center of the setup at a fixed position.
Steady-state temperatures are $\SI{69}{\kelvin}$ at the heat shield and $\SI{4.1}{\kelvin}$ at the SCC.
Due to the very high electrical conductivity of OFHC copper at such temperatures, RF magnetic-field noise at the trap is attenuated by eddy currents in the heat shields.
Measured in the similar setup at PTB,\cite{leopold_cryogenic_2019} the suppression is $30$ to $\SI{40}{\decibel}$ between $\SI{60}{\hertz}$ and $\SI{1}{\kilo\hertz}$, with a low-pass cut-off frequency around $\SI{0.1}{\hertz}$, depending on the spatial orientation of the magnetic-field vector.
Here, the SCC enclosing the trapping region additionally shields slowly changing, quasi-static magnetic perturbations by the Meissner-Ochsenfeld effect.\cite{meissner_neuer_1933}

Twelve ports in the horizontal plane provide optical access to the trap center (R1), for instance, for the lasers used for ${}^9\text{Be}$ photoionization and Doppler cooling of ${}^9\text{Be}^+$, as well as for the collimated ${}^9\text{Be}$ atomic beam produced by an oven connected to the trap chamber.
Two ports along the trap axis serve for injection and re-trapping of HCI from the EBIT and are equipped with electrostatic lenses inside the thermal stages and with mirror electrodes protruding into the monolithic tank (R2).
Fluorescence from the trap center is collected with a cryogenic optics system\cite{warnecke_2019, warnecke_imaging_2021}
consisting of seven lenses (UV fused silica and $\text{CaF}_2$) relaying an image of the ions through a $\SI{2}{\milli\metre}$ aperture at the outer temperature stage.
Here, an aspheric lens (UV fused silica) projects it through a vacuum window onto the detection system, consisting of an electron-multiplying charge-coupled device camera (Andor iXon Ultra 888) and a photomultiplier tube.
By adjusting the vertical position of this lens, the magnification can be set between $7.8$ and $20$.
Including absorption and reflection losses, the collection efficiency between $\SI{300}{\nano\metre}$ and $\SI{440}{\nano\metre}$ is greater than $1.1\%$, and approximately $2.17\%$ at the wavelength of the ${}^9\text{Be}^+$ Doppler-cooling transition at $\SI{313}{\nano\metre}$, assuming spherical emission of the ion.

Narrow stainless-steel tubes mounted on the cryogenic-shield apertures restrict the solid angle of room-temperature radiation visible to the ion to $0.084\%$ of $4\pi$, similar to our earlier cryogenic Paul traps.\cite{schmoger_kalte_2017,leopold_cryogenic_2019}
This also limits particle flux from room-temperature regions to the trap, lowering there the residual gas density, suppressing HCI losses by collisions and charge-exchange reactions, and thus extending their storage time.\cite{schmoger_kalte_2017, micke_closed-cycle_2019}

%%%%%%%%%%%%%%%%%%%%%%%%%%%%%%%%%%%%%%%%%%%%%%%%%%%%%%%%%%%%%%%%%%%%%%%%%%%%
\begin{figure}[t]
\centering
\includegraphics[width=\columnwidth]{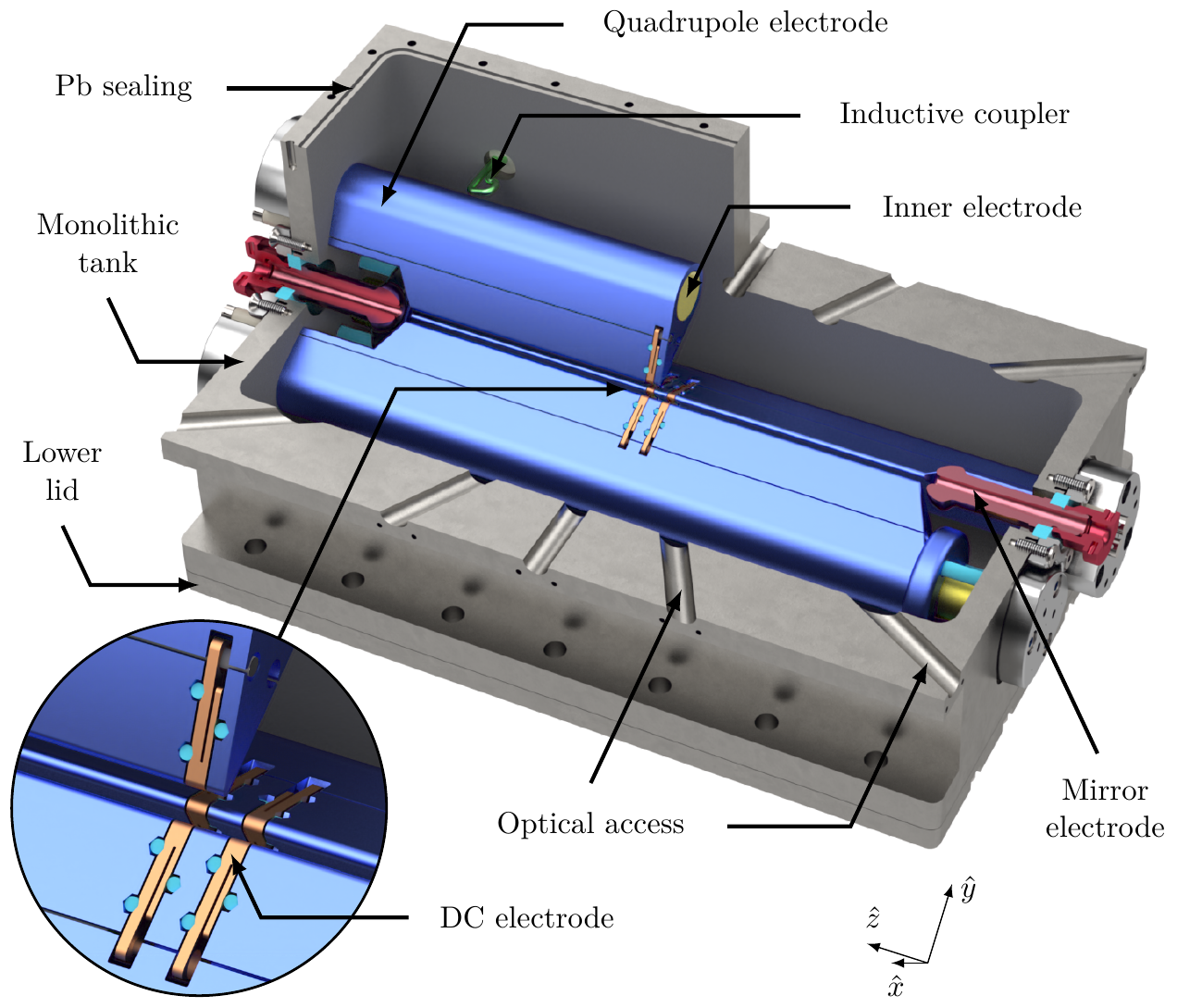}
\caption{Cutaway drawing of the niobium RF cavity.
Sapphire insulators are shown in cyan.
For details, see main text.
Adapted from Ref.\ \onlinecite{stark_phd_2020}.}
\label{Fig_CAD_model_SRF_cavity}
\end{figure}
%%%%%%%%%%%%%%%%%%%%%%%%%%%%%%%%%%%%%%%%%%%%%%%%%%%%%%%%%%%%%%%%%%%%%%%%%%%%

%%%%%%%%%%%%%%%%%%%%%%%%%%%%%%%%%%%%%%%%%%%%%%%%%%%%%%%%%%%%%%%%%%%%%%%%%%%%
\subsection{Superconducting cavity}
A CAD model of the RF cavity is shown in Fig \ref{Fig_CAD_model_SRF_cavity}.
The Paul trap quadrupole electrodes are an integral part of the resonator.
Its electric QM radially confines the ions as in a 2D-mass filter.
Biased direct current (DC) electrodes trap the ions along the symmetry axis of the quadrupole.
On its both ends, additional electrostatic mirror electrodes are used to capture injected HCI (R2).
All conducting parts are made of high-purity, massive niobium, a type-II superconductor with a critical temperature of $T_\text{c}=\SI{9.25}{\kelvin}$.\cite{Fischer2005}
For high mechanical stability (R3), the monolithic resonator tank supporting the quadrupole rods is machined from a single piece.
During cooldown, this suppresses differential contraction, which could lead to electrode misalignment.
Sapphire is used as insulator material: Its small dielectric loss at cryogenic temperatures\cite{krupka_complex_1999, tobar_proposal_2003} reduces RF power dissipation inside the SCC, and its high thermal conductivity of $\SI{230}{\watt\metre^{-1}\kelvin^{-1}}$ at $\SI{4}{\kelvin}$\cite{ekin_experimental_2006} improves thermalization of electrodes and tank.

%%%%%%%%%%%%%%%%%%%%%%%%%%%%%%%%%%%%%%%%%%%%%%%%%%%%%%%%%%%%%%%%%%%%%%%%%%%%
\subsubsection{Superconducting cavity tank and optical access}
The box-shaped cavity ($220 \times 140 \times 114 \,\text{mm}^3$) consists of the monolithic tank holding the coaxial quadrupole electrodes, the DC electrodes, and the electrostatic mirrors, as well as the top and bottom lids sealing it.
The upper lid holds a chevron-shaped disk made of Nb that gives optical access to the imaging system (see Fig.\ \ref{Fig_CAD_model_SRF_cavity_radial_section} and Fig.\ \ref{Fig_CAD_model_Cold_stage}) while suppressing RF emission, and thus cavity losses.
Its concentric rings and spokes transmit $83.2\%$ of the light from the trap center that is emitted within a solid angle of $\Omega/4\pi \simeq 0.106$.
A superconducting connection between these parts is established with high-purity $99.99\%$ lead wire with $T_\text{c}=\SI{7.2}{\kelvin}$.\cite{Fischer2005}
Twelve narrow bores through the side walls of the tank give access to the trap center in the horizontal plane (R1).
For suppressing RF leakage from the cavity, their diameter of $\SI{7}{\milli\metre}$, or $\SI{3.5}{\milli\metre}$ for the two axial ports, is chosen much smaller than the wavelength of the resonant mode, $\lambda_0 \approx \SI{8.7}{\metre}$.

%%%%%%%%%%%%%%%%%%%%%%%%%%%%%%%%%%%%%%%%%%%%%%%%%%%%%%%%%%%%%%%%%%%%%%%%%%%%
\subsubsection{Quadrupole electrodes}
A critical cavity design parameter is the resonance frequency $\omega_0$ of its electric QM corresponding to the drive frequency of the trap.
In a linear Paul trap, the stability parameter for the radial motion of an ion with charge $q$ and mass $m$ is given by
\begin{eqnarray}
|q_{r}| = \frac{4 q V_\text{RF}}{m r_0^2 \omega_0 ^2},
\label{Eq_Stability_Condition_Paul_trap}
\end{eqnarray}
where $V_\text{RF}$ is the RF voltage amplitude.
It defines the radial secular frequency $\omega_r \simeq q_r \omega_0 /\sqrt{8}$.
For efficient ground-state cooling of ${}^9\text{Be}^+$ ions, Lamb-Dicke parameters well below 1 and thus high secular frequencies on the order of $\omega_r/2\pi \simeq \SI{1}{\mega\hertz}$ are needed.\cite{leopold_cryogenic_2019}
Since the maximum voltage $V_\text{RF}$ is technically limited, one obtains an upper bound on the resonance frequency of about $\SI{100}{\mega\hertz}$.

%%%%%%%%%%%%%%%%%%%%%%%%%%%%%%%%%%%%%%%%%%%%%%%%%%%%%%%%%%%%%%%%%%%%%%%%%%%%
\begin{figure}[t]
\centering
\includegraphics[width=\columnwidth]{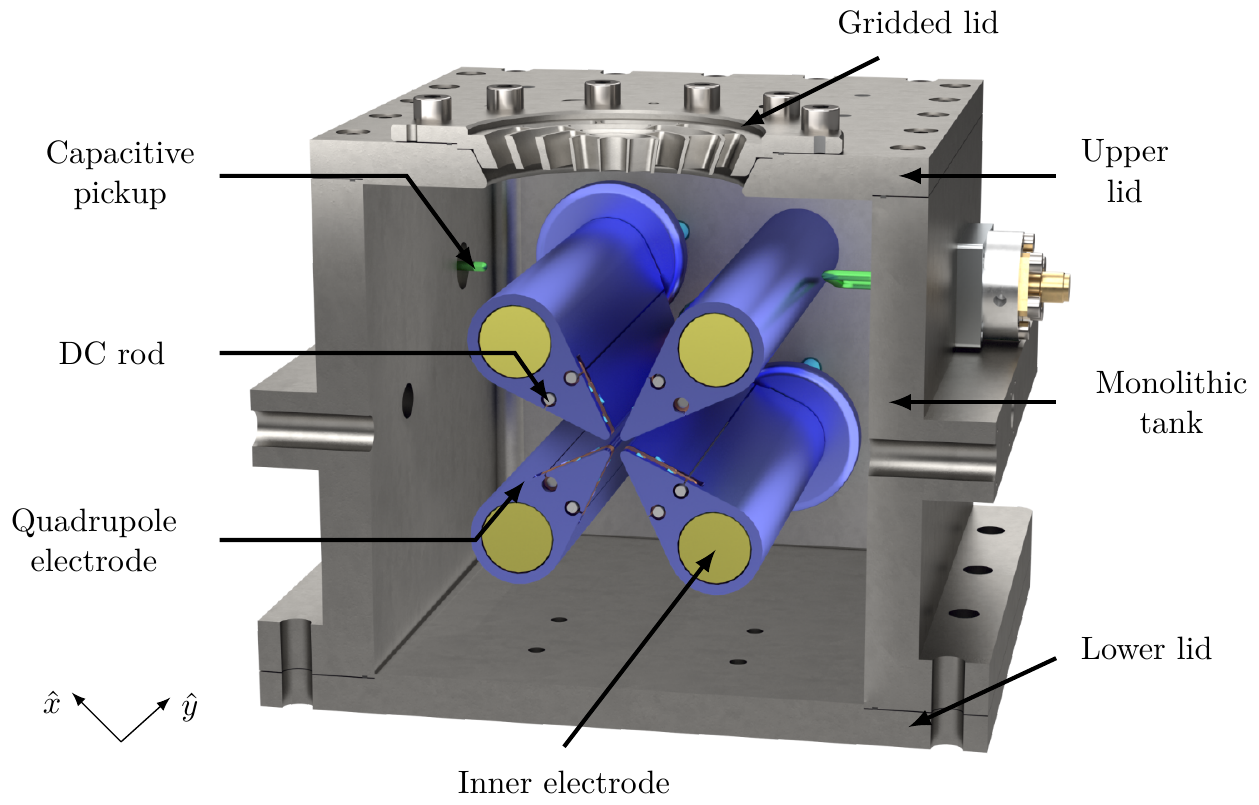}
\caption{Radial section of the RF cavity through the trap center at $z=0$.
Adapted from Ref.\ \onlinecite{stark_phd_2020}.}
\label{Fig_CAD_model_SRF_cavity_radial_section}
\end{figure}
%%%%%%%%%%%%%%%%%%%%%%%%%%%%%%%%%%%%%%%%%%%%%%%%%%%%%%%%%%%%%%%%%%%%%%%%%%%%

%%%%%%%%%%%%%%%%%%%%%%%%%%%%%%%%%%%%%%%%%%%%%%%%%%%%%%%%%%%%%%%%%%%%%%%%%%%%
\begin{figure}[b]
\centering
\includegraphics[width=\columnwidth]{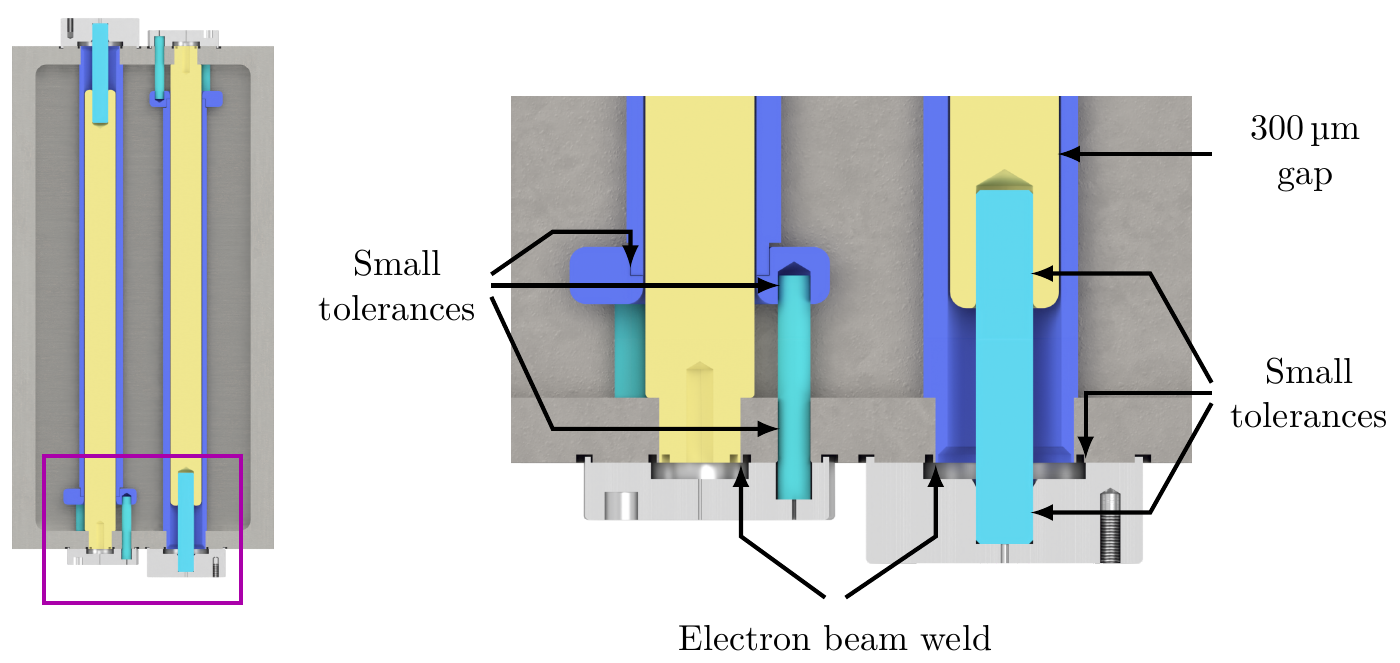}
\caption{Cut through the coaxial quadrupole electrodes, with inner (yellow) and outer (blue) segments separated by a $\SI{300}{\micro\metre}$ gap.
The concentric elements are electron-beam welded on one end to the tank and centered at the other one with sapphire insulators (cyan).
Reprinted from Ref.\ \onlinecite{stark_phd_2020}.}
\label{Fig_detailed_view_coaxial_electrodes}
\end{figure}
%%%%%%%%%%%%%%%%%%%%%%%%%%%%%%%%%%%%%%%%%%%%%%%%%%%%%%%%%%%%%%%%%%%%%%%%%%%%

We introduce a coaxial geometry illustrated in Fig.\ \ref{Fig_LC_Schematic_Working_Principle}, with each electrode having an inner and an outer conductor as opposite poles of the cavity QM.
Each of these conductors is electron-beam welded on one end to the tank wall (see Fig.\ \ref{Fig_detailed_view_coaxial_electrodes}), maintaining a superconducting connection, while the opposite end is centered by sapphire insulators.
Between them, a small gap of $\SI{300}{\micro\metre}$ generates a capacitance of $\sim \SI{230}{\pico\farad}$ in each electrode, resulting in a total quadrupole capacitance of $C_\text{QP}\approx \SI{928}{\pico\farad}$, which is two orders of magnitude higher than with single-rod electrodes.
This lowers the QM resonance frequency $\omega_0 \propto C_\text{QP}^{-1/2}$ into the desired range.
We designed the electrodes and the cavity with finite-element method (FEM) simulations discussed in Sec.\ \ref{Sec_FEM_simulations_resonance_frequency}.

To achieve a larger solid angle of the trap region towards the imaging system, the hyperbolic electrode geometry of an ideal Paul trap was sacrificed and a blade-style geometry chosen instead (see Fig.\ \ref{Fig_CAD_model_SRF_cavity_radial_section}).\cite{gulde_experimental_2003}
The tapered section of the quadrupole electrodes has a tip radius of $r_e = \SI{0.9}{\milli\metre}$.
Anharmonicities of the radial potentials were reduced by optimizing the electrode geometry with FEM simulations (see Sec.\ \ref{Sec_FEM_Trapping_Potentials}), reaching a relative contribution of next-to-leading order terms\cite{deng_design_2015} to the quadrupole potential below $\SI{8e-7}{}$ for ion crystals with a radial diameter $<\SI{200}{\micro\metre}$.

%%%%%%%%%%%%%%%%%%%%%%%%%%%%%%%%%%%%%%%%%%%%%%%%%%%%%%%%%%%%%%%%%%%%%%%%%%%%
\subsubsection{DC electrodes}
\begin{SCfigure}
\centering
\includegraphics[width=0.57\columnwidth]{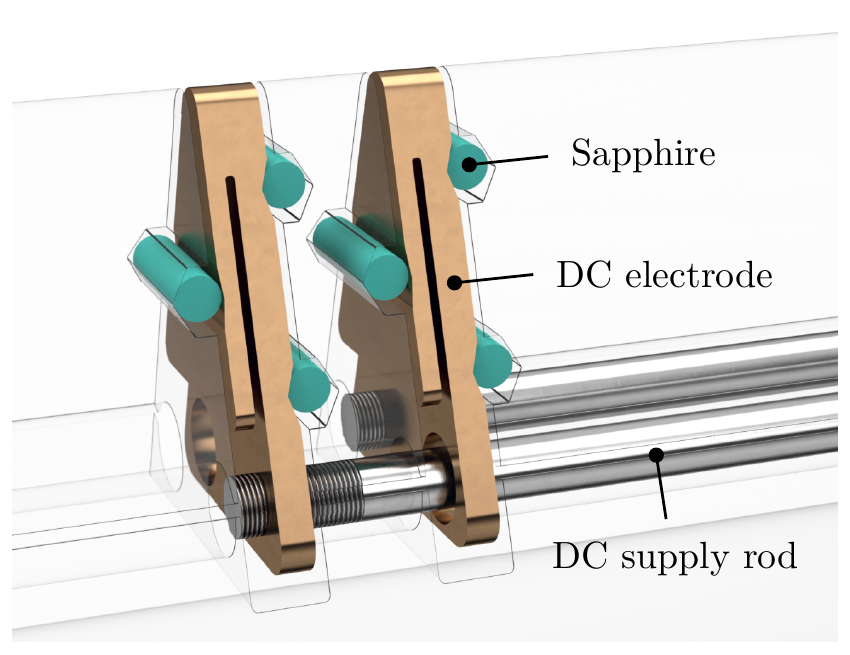}
\caption{Detailed view of the DC electrodes (orange) embedded within the quadrupole electrodes (only contours indicated) and held by three sapphire rods (cyan).
Adapted from Ref.\ \onlinecite{stark_phd_2020}.
}
\label{Fig_detailed_view_DC_electrodes}
\end{SCfigure}

For axial confinement of ions, eight DC electrodes are embedded inside the RF electrode structure, two of which are symmetrically integrated in each quadrupole electrode around the trap center, as can be seen in Fig.\ \ref{Fig_CAD_model_SRF_cavity} with a detailed view in Fig.\ \ref{Fig_detailed_view_DC_electrodes}.
The sliced electrodes are mounted under pre-tension and fixed in position using three sapphire rods of $\SI{1.5}{\milli\metre}$ diameter, providing electrical insulation and proper thermal contact, and a DC supply rod each.
The contour of the DC electrodes matches the tapered shape of the quadrupole electrodes and hides the sapphire insulators from the ions' direct line of sight by a judicious choice of their length.
This avoids stray electric fields on the trap axis due to insulator charge-up, e.g., by incident electrons, ions or UV photons.

The DC electrodes can be individually biased to generate the electrostatic potential required for axial ion confinement while minimizing excess micromotion that can arise from mismatching DC and RF nodes.\cite{berkeland_minimization_1998}
Electrical connections are provided by long niobium rods of $\SI{1.8}{\milli\metre}$ diameter, which are screwed into an $\text{M}1.6$ thread in the electrode.
Each rod leaves the cavity through a $\SI{3}{\milli\metre}$ bore beneath the surface of the respective quadrupole electrode (see Fig.\ \ref{Fig_detailed_view_DC_electrodes}).

Axial position and electrode geometry are both optimized for strong harmonic confinement by means of FEM simulations of the axial trapping potential, presented in Sec.\ \ref{Sec_FEM_Trapping_Potentials}.
The DC electrodes are axially separated by $2z_0= \SI{4.1}{\milli\metre}$. Within a distance of $|z|<\SI{250}{\micro\metre}$ from the minimum, this yields a relative contribution of anharmonic terms (up to the sixth order) to the axial potential below $2.9(1)\times10^{-5}$.

%%%%%%%%%%%%%%%%%%%%%%%%%%%%%%%%%%%%%%%%%%%%%%%%%%%%%%%%%%%%%%%%%%%%%%%%%%%%
%%%%%%%%%%%%%%%%%%%%%%%%%%%%%%%%%%%%%%%%%%%%%%%%%%%%%%%%%%%%%%%%%%%%%%%%%%%%

%%%%%%%%%%%%%%%%%%%%%%%%%%%%%%%%%%%%%%%%%%%%%%%%%%%%%%%%%%%%%%%%%%%%%%%%%%%%
\subsubsection{Mirror electrodes}
Retrapping of HCI inside the Paul trap will be implemented identically to the schemes described in Refs.\ \onlinecite{micke_coherent_2020} and \onlinecite{schmoger_deceleration_2015}.
The kinetic energy of the injected HCI is reduced by sequential Coulomb collisions with laser-cooled ${}^9\text{Be}^+$ ions prepared beforehand in the Paul trap.
Depending on the initial kinetic energy and the dimensions of the ${}^9\text{Be}^+$ crystal, thermalization and subsequent co-crystallization of the HCI requires about $10^2 - 10^3$ transits through the trap center.
This is realized by multiple reflections from the mirror electrodes mounted at both ends of the quadrupole structure (see Fig.\ \ref{Fig_CAD_model_SRF_cavity}).
Their large axial separation of $\SI{155}{\milli\metre}$ allows the entire HCI bunch (length on the order of $\SI{10}{\milli\metre}$)\cite{micke_be_2020} to enter the trap before the mirror electrode used for injection is switched to a higher potential to close the trap.

%%%%%%%%%%%%%%%%%%%%%%%%%%%%%%%%%%%%%%%%%%%%%%%%%%%%%%%%%%%%%%%%%%%%%%%%%%
\subsubsection{RF coupling}
\begin{figure}[t]
\centering
\includegraphics[width=\columnwidth]{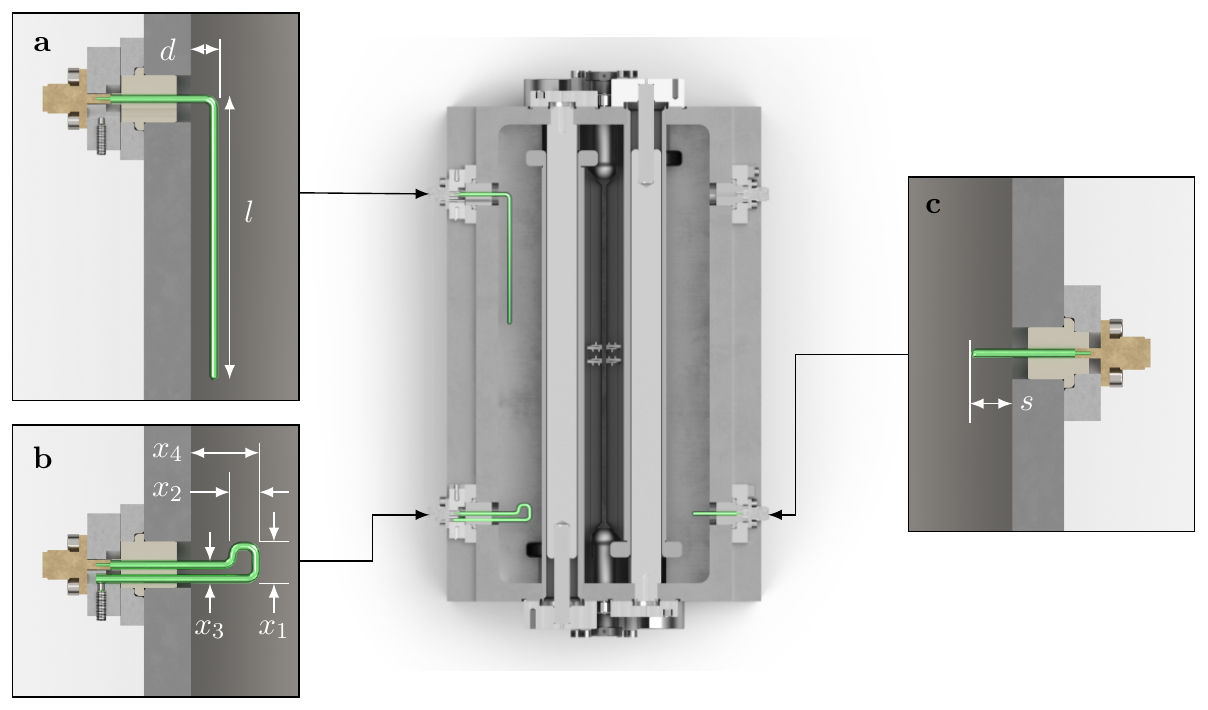}
\caption{Horizontal section through the RF cavity at the height of the upper coaxial electrodes.
The detail views show the RF couplers made of $\SI{2}{\milli\metre}$ thick niobium wire (green), which are soldered into SMA connectors (gold).
The couplers are isolated from their mounts by PTFE sleeves (beige) and fixed to the cavity tank with aluminium holders (light grey).
(\textbf{a}) Microwave antenna: $l=\SI{57}{\milli\metre}$, $d=\SI{5.9}{\milli\metre}$.
It is tilted by $\sim 17^\circ$ into the plane (not shown).
(\textbf{b}) Inductive coupling to the magnetic field: $x_1=\SI{9}{\milli\metre}$, $x_2=\SI{7}{\milli\metre}$, $x_3=\SI{5}{\milli\metre}$, and $x_4=\SI{14.9}{\milli\metre}$.
(\textbf{c}) Capacitive coupling to the electric field: $s=\SI{4.9}{\milli\metre}$.
Adapted from Ref.\ \onlinecite{stark_phd_2020}.}
\label{Fig_RF_Coupler}
\end{figure}

Three types of RF couplers are installed at the cavity, as can be seen in Fig.\ \ref{Fig_RF_Coupler}.
%%%%%%%%%%%%%%%%%%%%%%%%%%%%%%%%%%%%%%%%%%%%%%%%%%%%%%%%%%%%%%%%%%%%%%%%%%%%
%%%%%%%%%%%%%%%%%%%%%%%%%%%%%%%%%%%%%%%%%%%%%%%%%%%%%%%%%%%%%%%%%%%%%%%%%%%%
Coupling to its QM is realized using one capacitive pickup, which couples to the electric field, and one inductive loop coupling to the magnetic field.
The latter is used for the in-coupling of RF power during operation of the SCC.
One end of the loop is RF-grounded to the cavity-tank wall.
Reflected power from the loop is minimized by matching its impedance to the RF source.
For this, we adjust its angle with respect to the magnetic flux direction of the resonant mode.

For monitoring the electromagnetic field inside the cavity we use a capacitive probe, which is weakly coupled to the cavity.
It can also be used to stabilize the RF-drive frequency with respect to the quadrupole resonance.\cite{lindstrom_pound-locking_2011, pound_electronic_1946, drever_laser_1983}

In addition, a microwave $\lambda/4$ antenna for driving the ${}^2\text{S}_{1/2} (F=2)\rightarrow {}^2\text{S}_{1/2} (F=1)$ hyperfine transition in ${}^9\text{Be}^+$  at $\SI{1250}{\mega\hertz}$ is installed inside the cavity.\cite{leopold_cryogenic_2019}
It is mounted at an oblique angle to all trap axes as well as the external quantization axis, thus coupling to all Zeeman components of the transition.
For the commissioning measurements presented below, the microwave antenna was replaced by a second capacitive coupler.

%%%%%%%%%%%%%%%%%%%%%%%%%%%%%%%%%%%%%%%%%%%%%%%%%%%%%%%%%%%%%%%%%%%%%%%%%%%%
\subsection{Electronics}\label{Sec_Design_Electronics}
%%%%%%%%%%%%%%%%%%%%%%%%%%%%%%%%%%%%%%%%%%%%%%%%%%%%%%%%%%%%%%%%%%%%%%%%%%%%
\begin{figure}[t]
\centering
\includegraphics[width=\columnwidth]{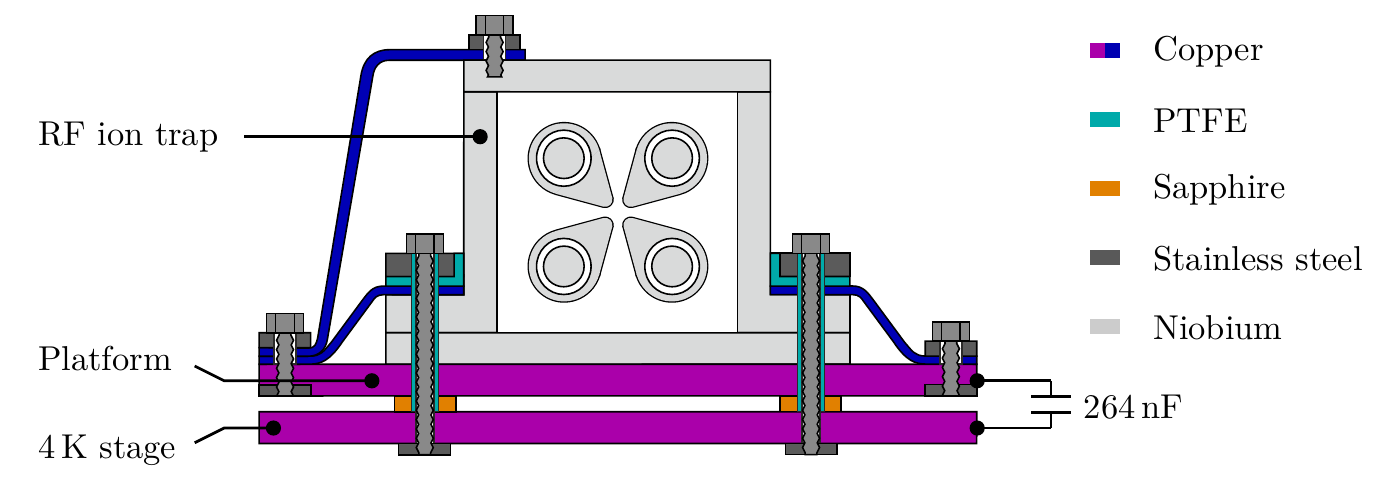}
\caption{Schematic drawing of the RF ion trap on the biased platform at $\SI{4}{\kelvin}$.
All parts of the cavity tank are individually thermalized using solid copper links (blue).
Reprinted from Ref.\ \onlinecite{stark_phd_2020}.}
\label{Fig_El_GND_Schematic}
\end{figure}
%%%%%%%%%%%%%%%%%%%%%%%%%%%%%%%%%%%%%%%%%%%%%%%%%%%%%%%%%%%%%%%%%%%%%%%%%%%%
The RF cavity rests on a copper platform, which is electrically isolated from its $\SI{4}{\kelvin}$ stage, as shown in Fig.\ \ref{Fig_El_GND_Schematic}, and can be biased for the deceleration of incoming HCI. It is connected for RF grounding to the $\SI{4}{\kelvin}$ stage by a capacitor set of $\SI{264}{\nano\farad}$, corresponding to an impedance of $\SI{18}{\milli\ohm}$ at the quadrupole resonance.
All electrical connections to the trap have a length of $\SI{2}{\metre}$ between their respective room-temperature vacuum feedthroughs and $\SI{4}{\kelvin}$ connectors.
The wires are thermally anchored at both cryogenic stages with $\SI{1}{\metre}$ length in between to reduce thermal conduction.
Electrical connections to the RF antennas use semi-rigid beryllium-copper coaxial cables (Coax Co., SC-219/50-SB-B) with PTFE dielectric.
All DC connections employ Kapton-insulated $\SI{200}{\micro\metre}$ thick phosphor-bronze wires.

\begin{figure}[b]
\centering
\includegraphics[width=\columnwidth]{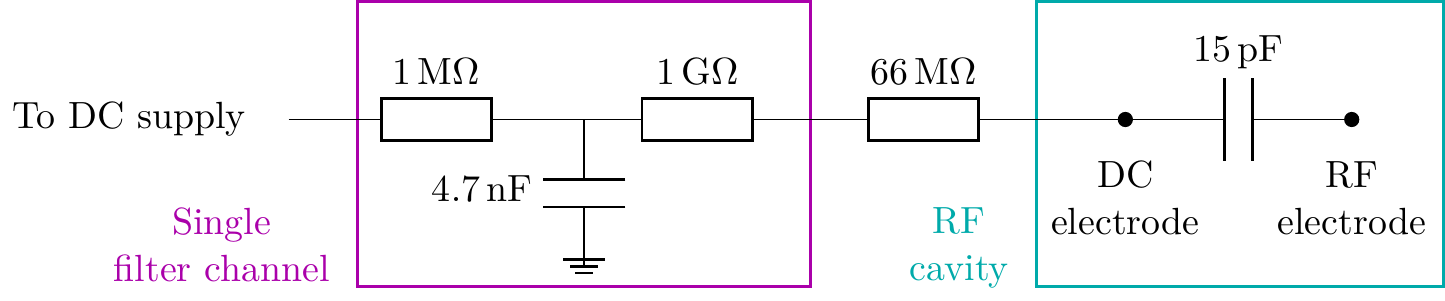}
\caption{Schematic of the effective $\SI{4}{\kelvin}$ electronic filter circuit connecting a single DC electrode to the external power supply. Surface-mounted components are soldered onto a Rogers RO4350B board grounded to the cavity tank.
For details, see main text.
Reprinted from Ref.\ \onlinecite{stark_phd_2020}.}
\label{Fig_4K_Electronics_Schematic}
\end{figure}

We use two-way, single-stage low-pass filters (shown in Fig.\ \ref{Fig_4K_Electronics_Schematic}), identical to the ones described in Ref.\ \onlinecite{leopold_cryogenic_2019}, with cut-off frequencies of $\SI{30}{\hertz}$ and $\SI{30}{\milli\hertz}$, respectively, for signals to and from the DC electrodes.
They suppress noise at the DC electrodes while protecting the DC power supplies from RF pickup.

Due to the small distance between the surrounding quadrupole electrode and each DC electrode including its supply rod, a stray capacitance of about $\SI{15}{\pico\farad}$ (see Fig.\ \ref{Fig_detailed_view_DC_electrodes}) couples the quadrupole RF voltage to the DC electrodes.
In order to reduce the RF loss due to parasitic capacitive coupling of the DC wires to RF ground, each rod is connected to its filtered DC power supply via a $\SI{66}{\mega\ohm}$ resistor.
This lets the DC-biased electrodes also oscillate at the full RF potential, which ensures strong radial confinement of transiting HCI and thus efficient retrapping (R2).

%%%%%%%%%%%%%%%%%%%%%%%%%%%%%%%%%%%%%%%%%%%%%%%%%%%%%%%%%%%%%%%%%%%%%%%%%%%%
%%%%%%%%%%%%%%%%%%%%%%%%%%%%%%%%%%%%%%%%%%%%%%%%%%%%%%%%%%%%%%%%%%%%%%%%%%%%

%%%%%%%%%%%%%%%%%%%%%%%%%%%%%%%%%%%%%%%%%%%%%%%%%%%%%%%%%%%%%%%%%%%%%%%%%%%%
%%%%%%%%%%%%%%%%%%%%%%%%%%%%%%%%%%%%%%%%%%%%%%%%%%%%%%%%%%%%%%%%%%%%%%%%%%%%
\section{Cavity production}\label{sec:III}

%%%%%%%%%%%%%%%%%%%%%%%%%%%%%%%%%%%%%%%%%%%%%%%%%%%%%%%%%%%%%%%%%%%%%%%%%%%%
\subsection{Material selection}

We manufactured the cavity from a high-purity niobium ingot.
The desired quality factor of $Q_0 = 10^5 - 10^7$ at a maximum electric-field gradient of $\SI{6}{\mega \volt\metre^{-1}}$ is much lower than the specified values for state-of-the-art superconducting cavities employed at accelerator facilities as, for example, the European X-Ray Free-Electron Laser (EuXFEL).\cite{singer_production_2016,singer_superconducting_2015}
Thus, our acceptable impurity levels (Tab.\ \ref{SRF_cavity_table_RRR}) are higher than there.
%Since the desired quality factor of $Q_0 = 10^5 - 10^7$ at a maximum electric-field gradient of $\SI{6}{\mega \volt\metre^{-1}}$ is much lower than the specified values for the cavities of the European X-Ray Free-Electron Laser (EuXFEL),\cite{singer_production_2016,singer_superconducting_2015} our acceptable impurity levels (Tab.\ \ref{SRF_cavity_table_RRR}) are higher than there.
%%%%%%%%%%%%%%%%%%%%%%%%%%%%%%%%%%%%%%%%%%%%%%%%%%%%%%%%%%%%%%%%%%%%%%%%%%%%
\begin{table}[b]
\centering
\caption{Excerpt of the chemical analysis of the niobium ingot used in our cavity, as provided by the supplier.
The material with a total purity $>99.86\%$ was purchased from Companhia Brasileira de
Metalurgia e Minera\c{c}\~{a}o in Brazil.}
\begin{tabular}{lclc}
\hline
\hline
Impurity \hphantom{$\frac{\Delta\rho}{\Delta C}$} & $\Delta m/m\,(\textrm{ppm})$ & Impurity \hphantom{$\frac{\Delta\rho}{\Delta C}$} & $\Delta m/m\,(\textrm{ppm})$\\
\hline
Nitrogen	& 18	&	Zirconium 	& 1	\\
Oxygen	    & 4		&	Tungsten 	& 5	\\
Carbon   	& 30	&	Tantalum 	& 1233	\\
Titanium	& 7		&	Hydrogen 	& 2	\\
Hafnium 	& 2		&	Molybdenum 	& 2	\\
Nickel 		& 1		&	Iron 		& 3	\\
\hline
\hline
\end{tabular}
\label{SRF_cavity_table_RRR}
\end{table}
%%%%%%%%%%%%%%%%%%%%%%%%%%%%%%%%%%%%%%%%%%%%%%%%%%%%%%%%%%%%%%%%%%%%%%%%%%%%
In comparison with the EuXFEL cavities,\cite{singer_superconducting_2015}  the concentration limits for N, C (both $\leq 10\,\text{ppm}$) and Ta ($500\,\text{ppm}$) are exceeded.
A residual resistance ratio (RRR) was not specified by the supplier.
We obtain an upper limit for the RRR using the dependence of the Nb resistivity on the concentration of some impurities.\cite{schulze_preparation_1981}
The total effect (Tab.\ \ref{SRF_cavity_table_RRR}) is a residual resistivity of $\SI{5.7e-10}{\ohm\metre}$ at $\SI{4.2}{\kelvin}$, or an $\text{RRR}<267$, slightly lower than the $\text{RRR}>300$ specified for the EuXFEL.

%%%%%%%%%%%%%%%%%%%%%%%%%%%%%%%%%%%%%%%%%%%%%%%%%%%%%%%%%%%%%%%%%%%%%%%%%%%%
\subsection{Fabrication and surface preparation}

Mechanical machining of metallic surfaces can contaminate them, limiting the performance of Nb as an RF superconductor.\cite{kelly_surface_2017}
The thickness of this so-called damaged or dirty layer depends on the manufacturing process, and varies between $50$ and $\SI{200}{\micro\metre}$.\cite{kelly_surface_2017}
Therefore, we instead employed non-intrusive electrical discharge machining (EDM) by wire at small material ablation rates to suppress heating.
This avoids gettering of hydrogen, oxygen, and other gases by Nb at temperatures above $200\,^{\circ}\text{C}$.\cite{bauer_fabrication_1980}
All parts were cut from the ingot using wire EDM in several steps with decreasing material ablation rate, yielding a well-defined contour with a smooth surface. Subsequently, all threads, holes, grooves and fits were milled. After degreasing and cleaning, the cavity parts were sent to an external company for electropolishing. Due to our gentle manufacturing, only $\SI{50}{\micro\metre}$ had to be removed from all surfaces.
Finally, all parts went through several cycles of ultrasonic cleaning in isopropyl alcohol, ethanol, and distilled as well as de-ionized water.

\begin{figure}[t]
\centering
\includegraphics[width=\columnwidth]{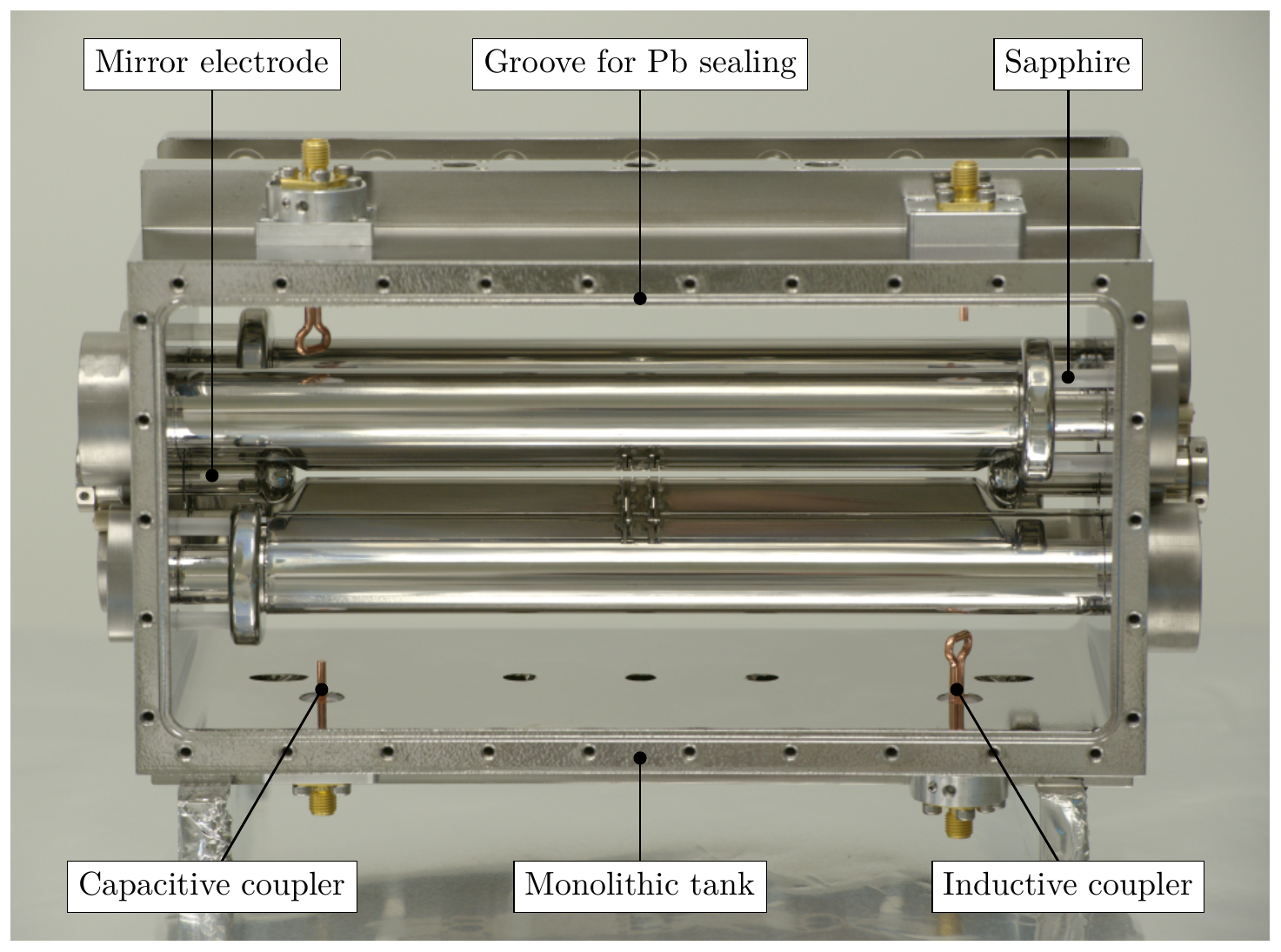}
\caption{Electropolished cavity with all lids removed.
For the measurements presented here, the temporary RF couplers made of copper that are shown in this image were later replaced by niobium parts and the second inductive coupler was removed.
%The image shows temporary RF couplers made of copper that were later replaced by niobium parts for the measurements presented here.
Adapted from Ref.\ \onlinecite{stark_phd_2020}.}
\label{Fig_Picture_SRF_Cavity}
\end{figure}

The cavity was assembled inside an ISO6 clean room to avoid surface contamination.
This was followed by electron-beam welding of cavity tank and electrodes performed at a background pressure $<\SI{1.2e-4}{\milli\bar}$. During this step, the cavity was kept closed except for thin tubes for evacuation of its interior.
A total of $\SI{7}{\hour}$ exposure to clean-room air between cleaning and welding of the cavity parts was not exceeded.
Finally, the DC supply rods as well as the RF couplers were installed, and the top and bottom lids were sealed using high-purity lead wire.
A picture of the fully-assembled cavity without the lids can be seen in Fig.\ \ref{Fig_Picture_SRF_Cavity}.
%%%%%%%%%%%%%%%%%%%%%%%%%%%%%%%%%%%%%%%%%%%%%%%%%%%%%%%%%%%%%%%%%%%%%%%%%%%%
%%%%%%%%%%%%%%%%%%%%%%%%%%%%%%%%%%%%%%%%%%%%%%%%%%%%%%%%%%%%%%%%%%%%%%%%%%%%

%%%%%%%%%%%%%%%%%%%%%%%%%%%%%%%%%%%%%%%%%%%%%%%%%%%%%%%%%%%%%%%%%%%%%%%%%%%%
%%%%%%%%%%%%%%%%%%%%%%%%%%%%%%%%%%%%%%%%%%%%%%%%%%%%%%%%%%%%%%%%%%%%%%%%%%%%
\section{FEM simulations}

\subsection{Simulation of the cavity resonant modes}\label{Sec_FEM_simulations_resonance_frequency}

We designed the cavity geometry for an electric QM with a resonance frequency on the order of $\SI{10}{\mega\hertz}$ by means of FEM simulations of the electromagnetic eigenmodes using the commercial COMSOL MULTIPHYSICS software and its RF module.

At the start of the simulation, an automatic mesh is generated, resolving narrow regions, e.g.\ the gaps between the coaxial electrodes.
On this mesh, the resonance frequencies and the eigenmodes are found by solving the wave equation on each mesh element with a plane-wave ansatz.
For nonlinear differential equations, a transformation point is chosen to linearize the functions around the given frequency.

The simulations were performed for a cavity at $T=\nobreak \SI{293.15}{\kelvin}$ in perfect vacuum and a perfectly conducting box as boundary condition.
All cavity elements were assigned identical material properties.
Lacking knowledge of the dielectric characteristics of the niobium material utilized for cavity fabrication, the relative permittivity and permeability were both set to unity.
The electrical conductivity was maximized within the restricted computational resources and is given by $\sigma=\SI{5.7e12}{\siemens\metre^{-1}}$.
The simulated eigenfrequencies of the four resonant modes between $\SI{1}{\mega\hertz}$ and $\SI{100}{\mega\hertz}$ are listed in Tab.\ \ref{Tab_SRF_cavity_summary_eigenfrequency_simulations}.
They are obtained with five significant digits and a relative simulation tolerance of $1\times10^{-5}$, which is $< \SI{1}{\kilo\hertz}$ for all resonant modes.
Within this accuracy, all eigenfrequencies exhibit a vanishing imaginary part that represents the internal RF losses of the geometry.

%%%%%%%%%%%%%%%%%%%%%%%%%%%%%%%%%%%%%%%%%%%%%%%%%%%%%%%%%%%%%%%%%%%%%%%%%%%%
\begin{table}[t]
	\centering
	\caption{Simulated eigenfrequencies of the four resonant modes between $\SI{1}{\mega\hertz}$ and $\SI{100}{\mega\hertz}$.
	The estimated uncertainties from mesh $\sigma_\text{m}$ and geometry $\sigma_\text{g}$ are added quadratically to yield the total uncertainty $\sigma_\text{tot}$.
	For details see text.}
	\begin{tabular}{cccc}
		\hline
		\hline
		$\omega_\text{0}/2\pi$ (MHz)& $\sigma_\text{m}$ (MHz) & $\sigma_\text{g}$ (MHz)&  $\sigma_\text{tot}$ (MHz)\\
		\hline
		$34.851$& $0.009$  		& $0.002$ 			&  $0.009$ \\
		$58.613$& $0.034$  		& $0.104$ 			&  $0.110$ \\
		$58.651$& $0.036$  		& $0.102$ 			&  $0.108$ \\
		$70.372$& $0.051$  		& $0.231$ 			&  $0.237$ \\
		\hline
		\hline
	\end{tabular}
	\label{Tab_SRF_cavity_summary_eigenfrequency_simulations}
\end{table}
%%%%%%%%%%%%%%%%%%%%%%%%%%%%%%%%%%%%%%%%%%%%%%%%%%%%%%%%%%%%%%%%%%%%%%%%%%%%

The simulations yield an electric quadrupole resonance at $\SI{34.851}{\mega\hertz}$.
Additional resonances at $\SI{58.613}{\mega\hertz}$ and $\SI{58.651}{\mega\hertz}$ exhibit a dipole-like structure of the radial electric field around the trap axis. At $\SI{70.372}{\mega\hertz}$, all outer quadrupole electrodes have the same RF potential.
Since the cavity confines the ions with the electric QM, excited with a narrowband RF signal, the other well-separated resonances are not further discussed.

%%%%%%%%%%%%%%%%%%%%%%%%%%%%%%%%%%%%%%%%%%%%%%%%%%%%%%%%%%%%%%%%%%%%%%%%%%%%
\subsubsection{Quadrupole resonant mode}
%%%%%%%%%%%%%%%%%%%%%%%%%%%%%%%%%%%%%%%%%%%%%%%%%%%%%%%%%%%%%%%%%%%%%%%%%%%%

\begin{figure*}[t]
\centering
\includegraphics[width=0.95\textwidth]{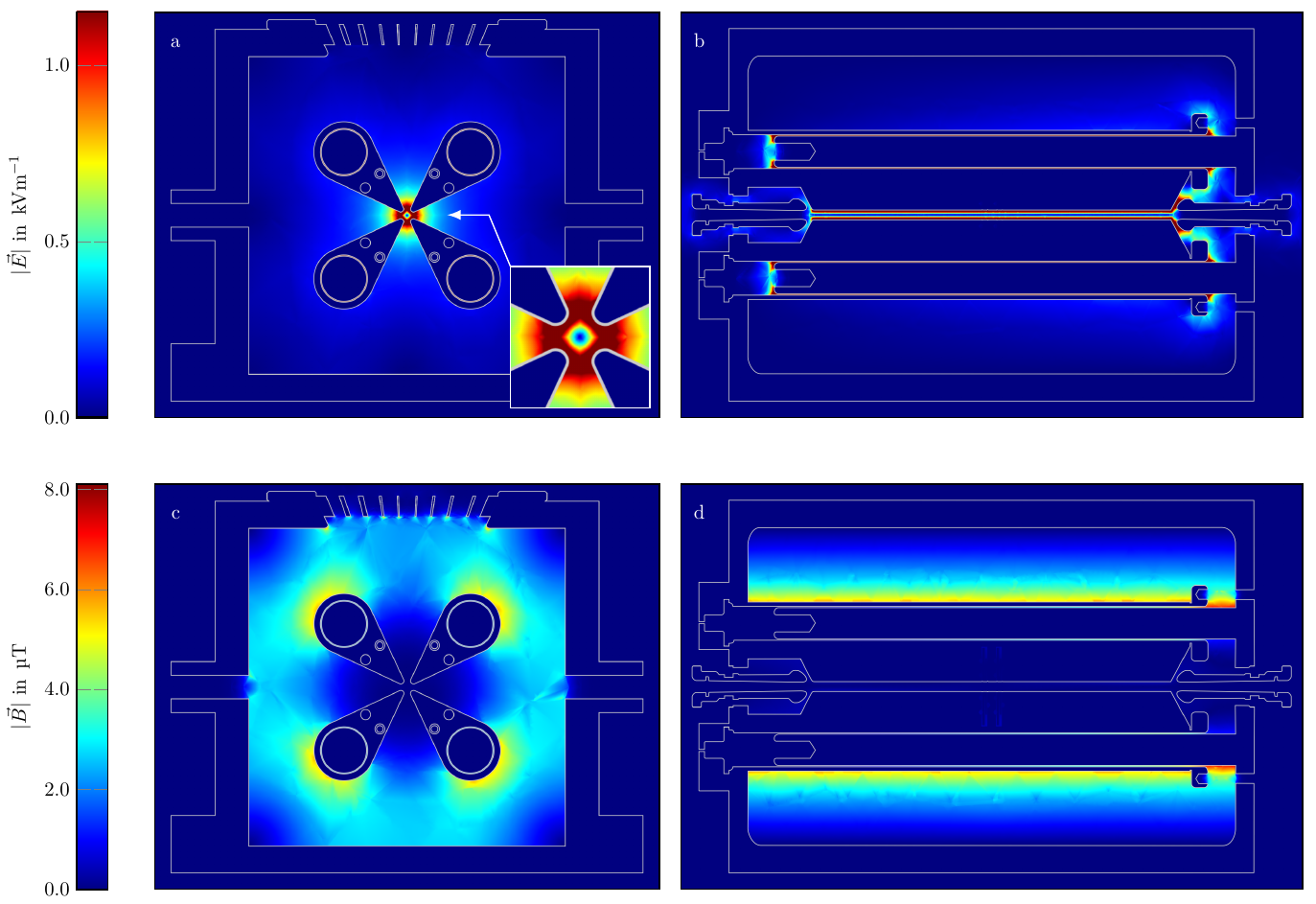}
\caption{FEM simulations of the RF electromagnetic field peak amplitudes of the QM.
\textbf{Top:} Electric field strength plotted in the radial plane at the axial trap center $z=0$ (\textbf{a}) and in the horizontal ($yz$)-plane through the trap center at $x=0$ (\textbf{b}).
Note that the scale is cut off below the maximum field strength of $\SI{10}{\kilo\volt\metre^{-1}}$ to emphasize the changes around the trap axis.
\textbf{Bottom:} Magnetic field strength in the radial plane at $z=0$ (\textbf{c}) and in the ($yz$)-plane at $x=0$ (\textbf{d}).
Adapted from Ref.\ \onlinecite{stark_phd_2020}.}
\label{Fig_E_and_B_field_simulations}
\end{figure*}
%%%%%%%%%%%%%%%%%%%%%%%%%%%%%%%%%%%%%%%%%%%%%%%%%%%%%%%%%%%%%%%%%%%%%%%%%%%%

The RF field amplitudes of the QM are shown in Fig.\ \ref{Fig_E_and_B_field_simulations}.
The electric field strength around the quadrupole electrodes has peak values in between the coaxial elements and close to the trap axis, decaying towards the cavity walls (Fig.\ \ref{Fig_E_and_B_field_simulations}(a)).
The outer coaxial electrodes shape the quadrupole electric field on the trap axis.
As explained above, the simulation shows nearly identical RF potentials for the DC electrodes and the quadrupole rods, with vanishing fields between them (Fig.\ \ref{Fig_E_and_B_field_simulations}(b)).
Along the trap axis, the homogeneous distribution of the radial electric field causes a constant radial confinement strength.

The RF magnetic field inside the cavity (Fig.\ \ref{Fig_E_and_B_field_simulations}(c,d)) is zero around the trap axis due to the geometry used, and its field lines, closed around the quadrupole structure, lie in the $(xy)$-plane.
The peak values of the RF magnetic field are radially localized around the quadrupole electrodes close to the regions with maximum current density on the cavity inner surfaces.

The simulations (Fig.\ \ref{Fig_E_and_B_field_simulations}(a,c)) show only small leakage of electromagnetic energy through the optical ports, as RF fields do not penetrate deep into those openings due to their comparatively large wavelengths.
The numerical accuracy of the simulations (five digits) does not allow determination of possible residual RF losses through those ports.

%%%%%%%%%%%%%%%%%%%%%%%%%%%%%%%%%%%%%%%%%%%%%%%%%%%%%%%%%%%%%%%%%%%%%%%%%%%%
\subsubsection{Simulation uncertainties}

Different systematic effects affect the simulation.
Its quality depends on the variable mesh element size approximating the real geometry.
The minimum size should be smaller than the tiniest structures, while the maximum should be smaller than the resonant mode wavelength.
Thus, the mesh was refined down to a minimum size of $\SI{51}{\micro\metre}$ for convergence.
For varying minimum element size in the region $\leq \SI{200}{\micro\metre}$, the simulation results show fluctuations on the level of kHz for the QM and $\SI{10}{\kilo\hertz}$ for the other modes.
The final result for each mode listed in Tab.\ \ref{Tab_SRF_cavity_summary_eigenfrequency_simulations} is the average of these values, and the largest deviation between any two frequencies is given as systematic uncertainty $\sigma_\text{m}$.

Since the required computation time increases drastically with simulation volume and mesh refinement, the complex cavity geometry had to be simplified.
Elements such as insulators, RF couplers, threads, and lids at the outer surface of the cavity housing and in low-RF field regions only have a minor effect for the QM, and were thus removed.
The influence of this simplification on the eigenfrequencies was estimated by comparing simulations with a minimum mesh size of $\SI{0.8}{\milli\metre}$ for both complete and simplified cavity geometry.
The frequency shift that appeared was used to estimate the systematic uncertainty $\sigma_\text{g}$ in Tab.\ \ref{Tab_SRF_cavity_summary_eigenfrequency_simulations}.

%%%%%%%%%%%%%%%%%%%%%%%%%%%%%%%%%%%%%%%%%%%%%%%%%%%%%%%%%%%%%%%%%%%%%%%%%%%%
\subsection{Simulation of the ion-trap potentials}\label{Sec_FEM_Trapping_Potentials}
To suppress higher-order contributions to the 3D harmonic trapping potential and optimize the geometries of DC and quadrupole electrodes, we performed electrostatic FEM simulations with the COMSOL AC/DC module in several steps.
First, a mesh is generated with more detail close to the trap center and a simplified geometry stripped of insulators, DC supply rods, and inner coaxial electrodes and the cavity tank replaced by a cuboid representing its inner walls.
All surfaces are modeled as perfect electric conductors, and Dirichlet boundary conditions are applied.
The potential distribution within the simulation volume is then obtained with a relative simulation tolerance of $1\times10^{-10}$ by solving Gauss' law.

%%%%%%%%%%%%%%%%%%%%%%%%%%%%%%%%%%%%%%%%%%%%%%%%%%%%%%%%%%%%%%%%%%%%%%%%%%%%
\subsubsection{Axial confinement}
\begin{figure}[b]
\centering
\includegraphics[width=\columnwidth]{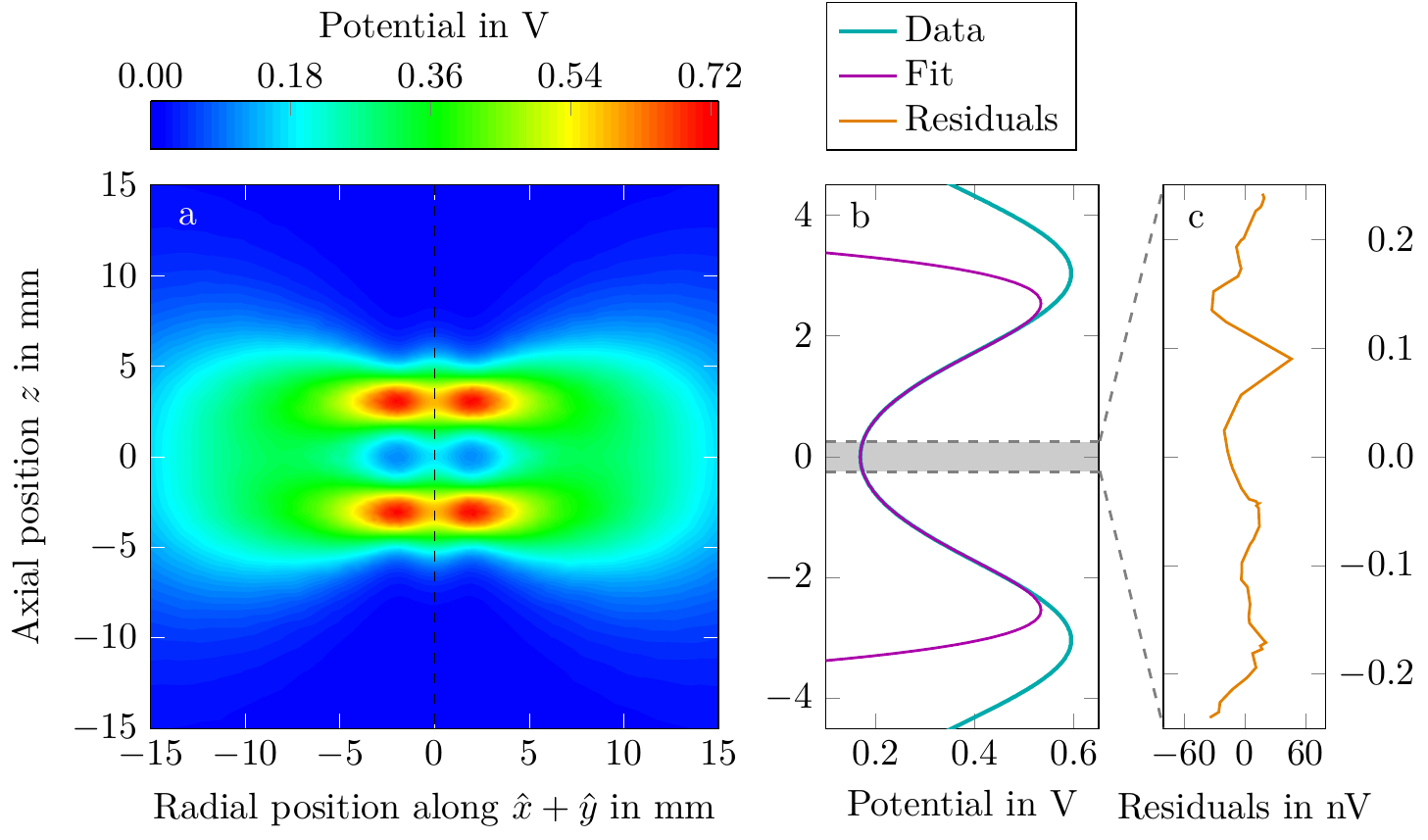}
\caption{Simulation of the DC electrode potential around the trap center with $\SI{1.0}{\volt}$ applied.
(\textbf{a}) 2D potential in the horizontal plane spanned by the trap axis $z$ (dashed line) and the direction $\hat{x}+\hat{y}$.
(\textbf{b}) Line-out along the trap axis with a sixth-order polynomial fit (see Eq.\ \ref{Eq_axial_polynomial_fit}) in the range of $\pm\SI{250}{\micro\metre}$ indicated by the gray region.
(\textbf{c}) Fit residuals (note the different scale).
Reprinted from Ref.\ \onlinecite{stark_phd_2020}.}
\label{Fig_FEM_ax_potentials}
\end{figure}

The simulated DC potential distribution in the horizontal plane around the trap center is shown in Fig.\ \ref{Fig_FEM_ax_potentials} for a minimum mesh element size of $\SI{10}{\micro\metre}$.
%%%%%%%%%%%%%%%%%%%%%%%%%%%%%%%%%%%%%%%%%%%%%%%%%%%%%%%%%%%%%%%%%%%%%%%%%%%%
%%%%%%%%%%%%%%%%%%%%%%%%%%%%%%%%%%%%%%%%%%%%%%%%%%%%%%%%%%%%%%%%%%%%%%%%%%%%
All elements of the simulation geometry are grounded, except the DC electrodes, which are biased to $\SI{1.0}{\volt}$.
The harmonicity of the axial potential is evaluated using the potential line-out along the trap axis ($z$), which can be represented as
\begin{eqnarray}
\Phi(z)=C_0 +C_2z^2+C_4z^4+C_6z^6+...
\label{Eq_axial_polynomial_fit}
\end{eqnarray}
around its minimum.
Fitting this model on different data ranges $\pm\Delta z$ yields the expansion coefficients up to $C_6$ listed in Tab.\ \ref{Tab_Fit_axial_potential}.
%%%%%%%%%%%%%%%%%%%%%%%%%%%%%%%%%%%%%%%%%%%%%%%%%%%%%%%
\begin{table}[t]
	\centering
	\caption{Coefficients of the sixth-order polynomial fit (see Eq.\ \ref{Eq_axial_polynomial_fit}) to the simulated trapping potential along the trap axis (see Fig.\ \ref{Fig_FEM_ax_potentials}) for a DC electrode bias of $U_\text{DC} = \SI{1.0}{\volt}$ at different fit ranges $\pm\Delta z$ around the potential minimum.
	The shift $\Delta \omega_z/\omega_z$ (see Eq.\ \ref{Eq_axial_frequency_shift}) of the axial secular frequency from listed anharmonicities is given for the maximum displacement $z$ within the fit range.}
		\begin{tabular}{cccc}
		\hline
		\hline
		\multicolumn{1}{r}{$\Delta z=$}  &  $0.25\,\text{mm}$ & $0.5\,\text{mm}$&  $1.0\,\text{mm}$\\
		\hline
		$C_2\,(10^{-2}\,\text{V}/\text{mm}^2)$ 	& $\,\,\,8.04159(9)$		& $\,\,\,\,\,\,8.04150(2)$ & $\,\,\,\,\,\,8.04129(2) $	\\
		$C_4\,(10^{-3}\,\text{V}/\text{mm}^4)$  	& $\,\,\,\,\,\,\,\,\,\,\,\,1.32(4)$		& $\,\,\,\,\,\,\,\,\,\,\,\,1.347 (2)$	& $\,\,\,\,\,\,\,\,\,1.3586 (5)$\\
		$C_6\,(10^{-4}\,\text{V}/\text{mm}^6)$ 	& $\,\,\,\,\,\,\,\,\,\,\,\,\,\,\,-8(5)$			& $\,\,\,\,\,\,\,\,\,\,- 8.53(4)$& $\,\,\,\,\,\,\,-8.727(4)$	\\
		\hline
		$\kappa=C_2 z_0^2/U_\text{DC}$						& $0.337948(4)$							& $0.3379439(6)$	& $0.3379351(8)$								\\
		$\Delta \omega_z/\omega_z\,(10^{-3})$ 		& $\,\,\,\,\,\,\,\,\,\,\,\,0.74(4)$		& $\,\,\,\,\,\,\,\,\,\,\,\,2.519(5)$	 & $\,\,\,\,\,\,\,\,\,\,\,\,2.497(6)$		\\
		\hline
		\hline
	\end{tabular}
	\label{Tab_Fit_axial_potential}
\end{table}
%%%%%%%%%%%%%%%%%%%%%%%%%%%%%%%%%%%%%%%%%%%%%%%%%%%%%%%%%%%%
Higher-order terms cause a dependence of the axial eigenfrequency of a trapped ion on its axial position, and thus on its energy.
The corresponding maximum frequency shift $\Delta \omega_z$ due to the first two anharmonicities, $C_4$ and $C_6$, can be calculated following Ref.\ \onlinecite{ulmer_first_2011}.
Using the approximation $\Delta \omega_z/\omega_z \ll 1$, it can be expressed as
\begin{eqnarray}
\frac{\Delta \omega_z}{\omega_z}=\frac{3}{4}\left( \frac{C_4}{C_2} +\frac{5}{4}\frac{C_6}{C_2} z^2  \right) z^2 .
\label{Eq_axial_frequency_shift}
\end{eqnarray}
The shifts at maximum ion displacement $z=\pm\Delta z$ from the potential minimum at $z=0$ are listed in Tab.\ \ref{Tab_Fit_axial_potential}.
For a single Doppler-cooled ${}^9\text{Be}^+$ ion ($T\approx\SI{300}{\micro\kelvin}$) with a secular-motion amplitude of $\SI{120}{\nano\metre}$ at $\omega_z /2\pi = \SI{1}{\mega\hertz}$, the anharmonicities of the smallest fit range translate to a maximum frequency shift of $\Delta\omega_z/\omega_z = 1.73(6) \times 10^{-10}$.

%%%%%%%%%%%%%%%%%%%%%%%%%%%%%%%%%%%%%%%%%%%%%%%%%%%%%%%%%%%%%%%%%%%%%%%%%%%%
\subsubsection{Radial confinement}

\begin{figure}[b]
\centering
\includegraphics[width=\columnwidth]{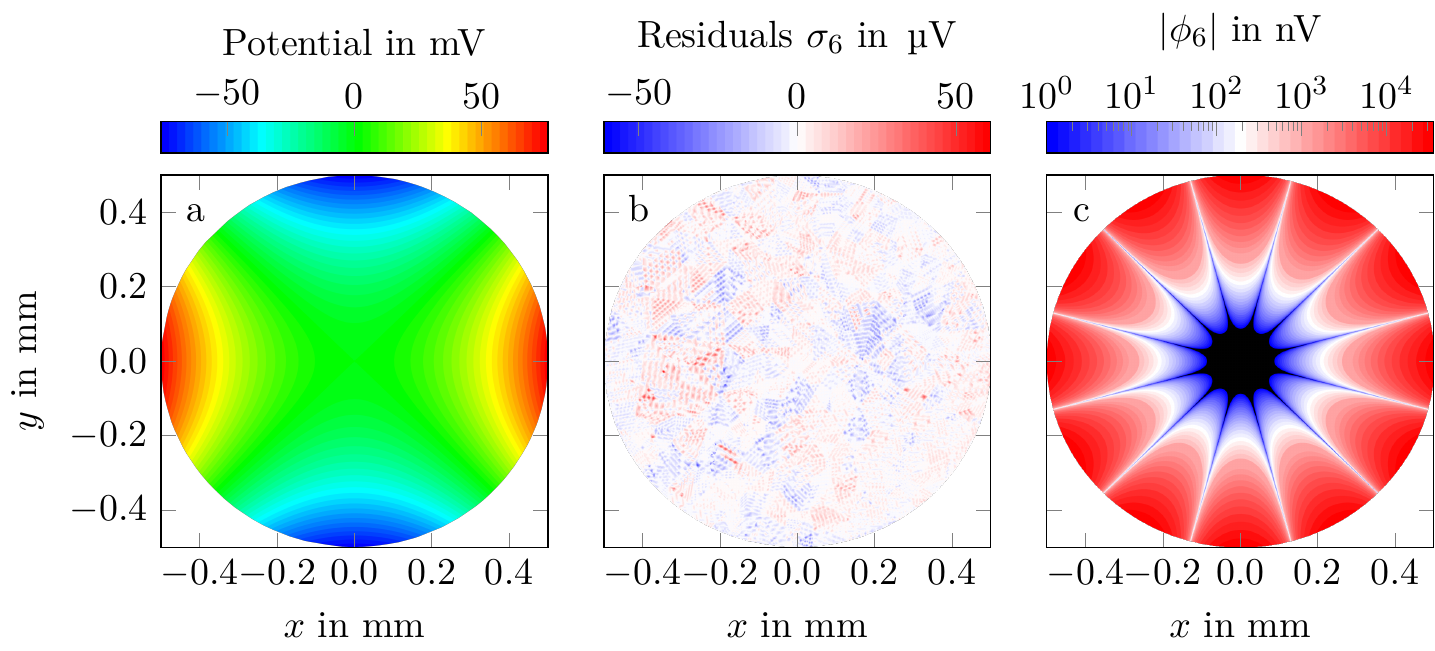}
\caption{Electrostatic simulation of the quadrupole-electrode potential at $z = 0$ for radial distances $r \leq \SI{0.5}{\milli\metre}$ to the trap axis.
(\textbf{a}) Simulated potential distribution for quadrupole potentials of $\pm\SI{1.0}{\volt}$.
(\textbf{b}) Residuals $\sigma_6$ of the sixth-order polynomial fit (see Eq.\ \ref{Eq_radial_polynomial_fit}).
(\textbf{c}) Absolute value of the first anharmonic contribution $\phi_6$ of the sixth-order polynomial fit.
Values $\leq \SI{1.0}{\nano\volt}$ are shown in black.
Reprinted from Ref.\ \onlinecite{stark_phd_2020}.}
\label{Fig_FEM_rad_potentials}
\end{figure}

The simulated quadrupole-electrode potential in the radial plane around the trap center at $z=0$ is shown in Fig.\ \ref{Fig_FEM_rad_potentials}(a) for a minimum mesh element size of $\SI{47.5}{\micro\metre}$.
%%%%%%%%%%%%%%%%%%%%%%%%%%%%%%%%%%%%%%%%%%%%%%%%%%%%%%%%%%%%%%%%%%%%%%%%%%%%
%%%%%%%%%%%%%%%%%%%%%%%%%%%%%%%%%%%%%%%%%%%%%%%%%%%%%%%%%%%%%%%%%%%%%%%%%%%%
Since quadrupole and DC electrodes are strongly RF coupled (see Sec.\ \ref{Sec_Design_Electronics}), all are set to common RF potentials of $\pm\SI{1.0}{\volt}$, while the remaining parts of the geometry are grounded.
Perpendicular to the symmetry axis of the quadrupole ($z$), the potential can be expanded in a multipole series:
\begin{eqnarray}
\Phi(x,y) = A_0 + \sum_{n} A_n \phi_{n}(x,y),
\label{Eq_radial_polynomial_fit}
\end{eqnarray}
where the first three terms are given by\cite{deng_design_2015}
\begin{alignat}{2}
\phi_{2} &= x^2-y^2, \nonumber\\
\phi_6 &= x^6-15x^4 y^2 +15 x^2y^4 -y^6,\nonumber\\
\phi_{10} &= x^{10}-45x^8y^2+210(x^6y^4-x^4y^6)+45x^2y^8-y^{10}.\nonumber
\end{alignat}
This two-dimensional model is fitted to the data for estimating deviations from the ideal quadrupole potential, $\phi_2$.
Results listed in Tab.\ \ref{Tab_Fit_radial_potential} show that only the next-to-leading order term $\phi_6$ contributes significantly.
%%%%%%%%%%%%%%%%%%%%%%%%%%%%%%%%%%%%%%%%%%%%%%%%%%%%%%%
\begin{table}[t]
	\centering
	\caption{Multipole-expansion coefficients of the quadrupole-electrode potential at $z=0$ for quadrupole potentials of $\pm \SI{1.0}{\volt}$.
	The parameters are determined by fitting a polynomial of $n^\text{th}$ order (see Eq.\ \ref{Eq_radial_polynomial_fit}) to the simulation data from Fig.\ \ref{Fig_FEM_rad_potentials}.
	The fit range is restricted to a radial distance of $r\leq\SI{0.5}{\milli\metre}$ to the trap axis.}
	\begin{tabular}{rccc}
		\hline
		\hline
		 $n=$ 	& $2$ & $6$ &  $10$\\
		\hline
		$A_2\,(10^{-1}\,\text{V}/\text{mm}^2)$		& $3.029238(5)$ 				& 	$3.029263(3)$								& 	$3.029263(3)$	\\
		$A_6\,(10^{-3}\,\text{V}/\text{mm}^6)$  	& 				& $\,\,\,\,\,\,2.2881(2)$			& 	$\,\,\,\,\,\,\,\,\,2.288(6)$\\
		$A_{10}\,(10^{-4}\,\text{V}/\text{mm}^{10})$  	& 				& 			& $\,\,\,\,\,\,-0.3 (1.1)$	\\
		\hline
		\hline
	\end{tabular}
	\label{Tab_Fit_radial_potential}
\end{table}
%%%%%%%%%%%%%%%%%%%%%%%%%%%%%%%%%%%%%%%%%%%%%%%%%%%%%%%%%%%%
Accordingly, the residuals of the sixth order polynomial fit (Fig.\ \ref{Fig_FEM_rad_potentials}(b)) do not show contributions from higher multipoles but rather reflect the mesh structure at this cut through the 3D simulation volume.
The absolute value of the anharmonicity $\phi_6$ of the radial potential is plotted in Fig.\ \ref{Fig_FEM_rad_potentials}(c).
Close to the trap axis ($r\leq \SI{100}{\micro\metre}$ for larger Coulomb crystals), its relative contribution to the radial potential is below $8\times 10^{-7}$.

%%%%%%%%%%%%%%%%%%%%%%%%%%%%%%%%%%%%%%%%%%%%%%%%%%%%%%%%%%%%%%%%%%%%%%%%%%%%
%%%%%%%%%%%%%%%%%%%%%%%%%%%%%%%%%%%%%%%%%%%%%%%%%%%%%%%%%%%%%%%%%%%%%%%%%%%%
\section{Resonant cavity quality factor}\label{Sec_Cavity_Characterization}
\begin{figure}[t]
\centering
\includegraphics[width=\columnwidth]{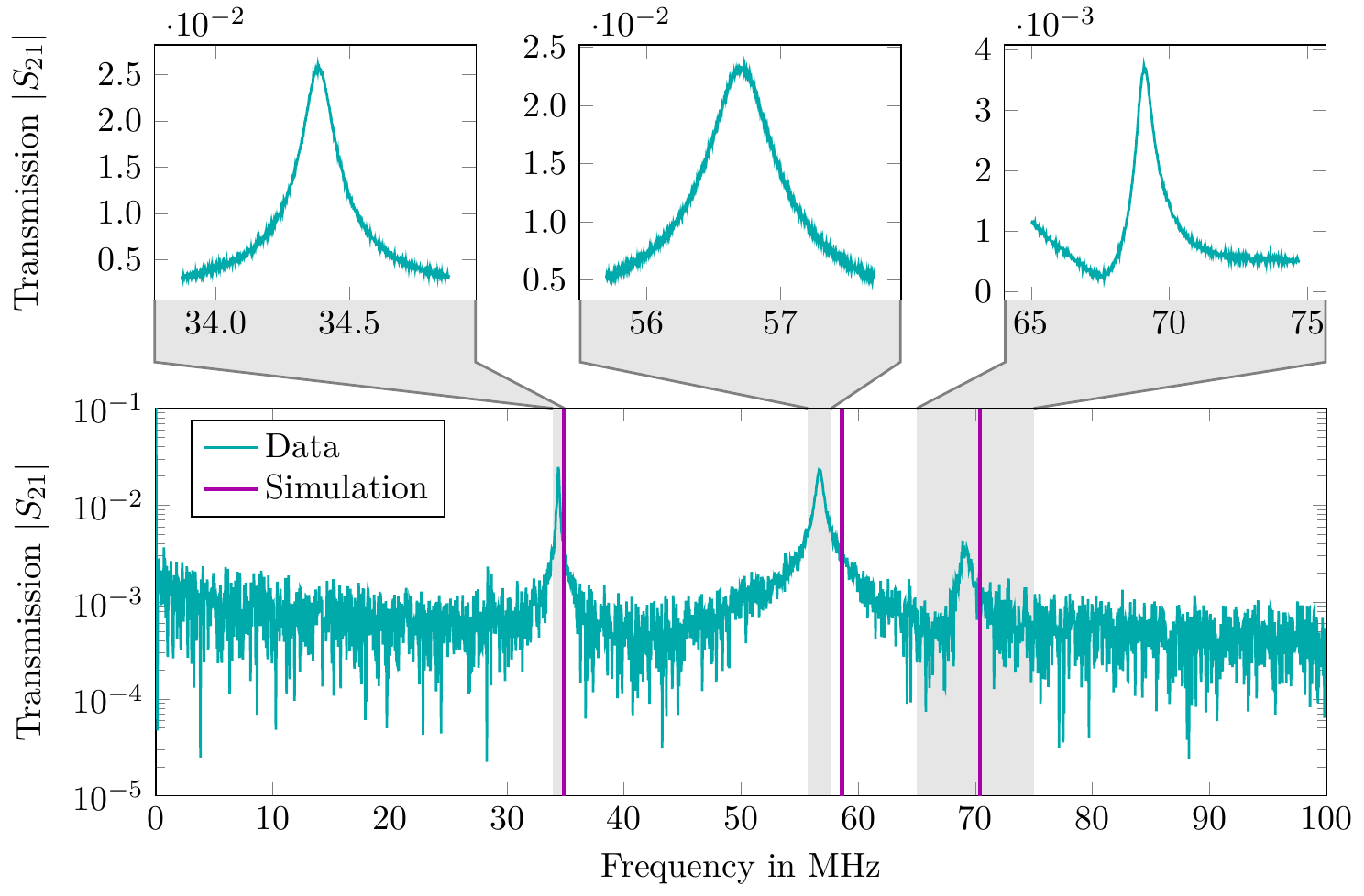}
\caption{Transmission response function of the cavity driven by the inductive coupler between $\SI{1}{\hertz}$ and $\SI{100}{\mega\hertz}$ at room temperature as recorded with the capacitive pick-up (cyan), and simulated eigenfrequencies from Sec.\ \ref{Sec_FEM_simulations_resonance_frequency} (magenta).
Reprinted from Ref.\ \onlinecite{stark_phd_2020}.}
\label{Fig_Transmission_exp_and_sim}
\end{figure}

\begin{figure}[t]
\centering
\includegraphics[width=\columnwidth]{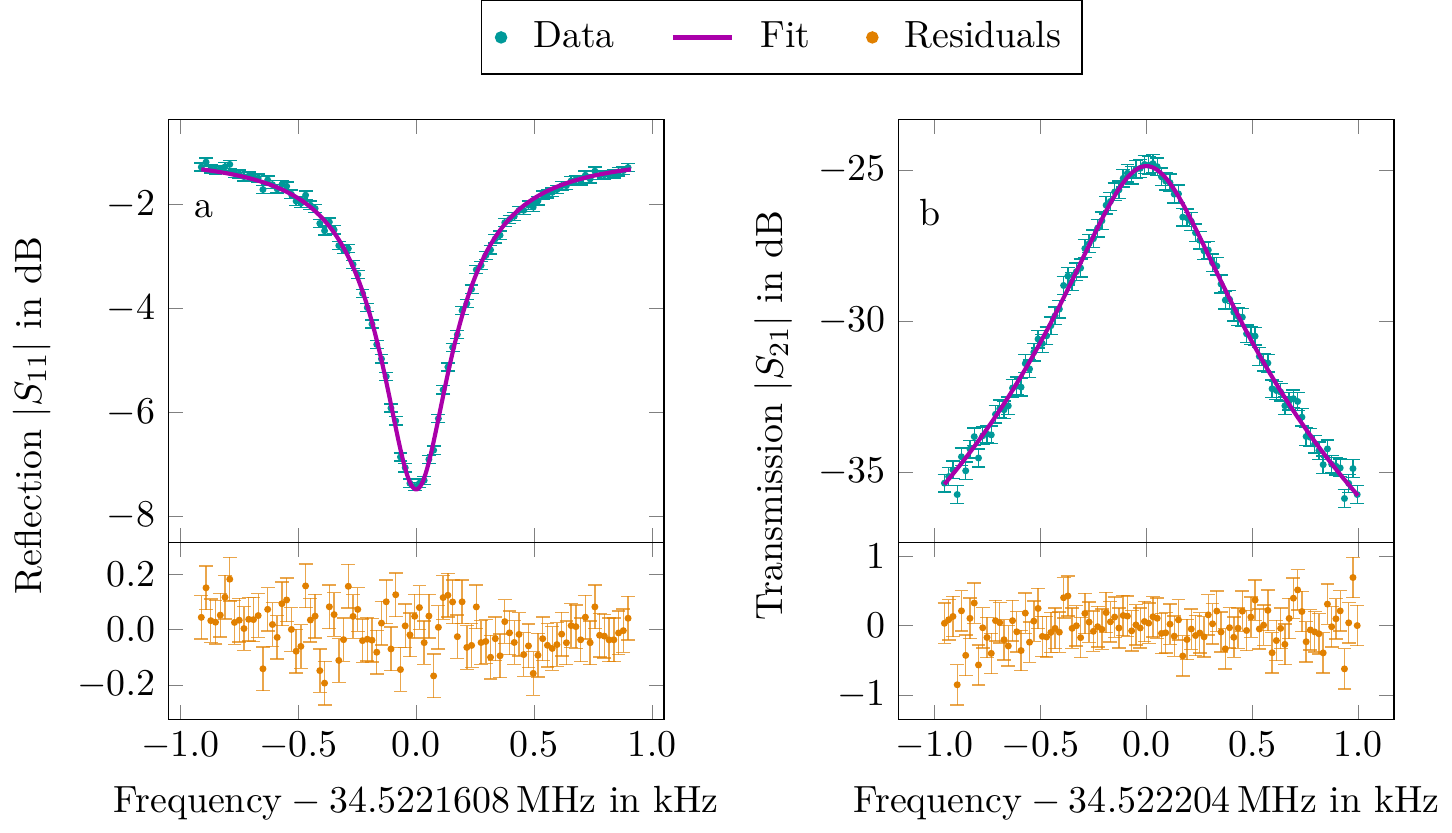}
\caption{Response function of the QM in reflection (\textbf{a}) and transmission (\textbf{b}) at $\SI{4.1}{\kelvin}$ temperature, with fits using Eq.\ \ref{eq_s_param_general} and respective residuals.
Adapted from Ref.\ \onlinecite{stark_phd_2020}.}
\label{Fig_Transmission_and_reflection_measured}
\end{figure}

We characterize the SCC by determining its QM quality factor at cryogenic temperatures.
Two common methods for this are cavity-ringdown and scattering-matrix measurements.
In the former approach, the decay time of stored energy following pulsed excitation is measured, yielding the quality factor of the corresponding resonance.
Here, we instead employ the second technique, based on the bandpass behaviour of near-resonant transmission and reflection spectra.
The power transmitted or reflected by the RF couplers in the cavity is described using the scattering-matrix ($\hat{S}$) formalism,\cite{caspers_rf_2012,dittes_decay_2000} giving the relation between the voltage amplitudes of incident ($a_i \sqrt{Z_0}$) and reflected ($b_i \sqrt{Z_0}$) waves at different ports:
\begin{eqnarray}
\vec{b}=\hat{S}\vec{a}\quad \text{with}\quad S_{ij}=\frac{b_i}{a_j},
\end{eqnarray}
with $Z_0$ being the transmission line impedance.
Each cavity port $i$ introduces losses parametrized by $P_{\text{d}, i}=\nobreak \omega_0 W {Q_i^{-1}}$ at resonance.
Including those, the loaded quality factor is given by $Q^{-1}=Q_0^{-1} + \sum_n Q_n^{-1}$,
where the unloaded quality factor $Q_0$ accounts only for cavity losses.
For an isolated resonance at $\omega_0$, the scattering between two ports $i$ and $j$ can be expressed as\cite{dittes_decay_2000}
\begin{eqnarray}
S_{ij}(\omega)= \delta_{ij}-\frac{2\sqrt{Q^2(Q_i Q_j)^{-1}}}{2\text{i} Q(\omega/\omega_0 -1)+1}.
\label{eq_s_param_general}
\end{eqnarray}
We carried out the presented measurements with a vector network analyzer (R\&S ZVL3) driving the cavity with the inductive coupler and probing it with the capacitive pickup.
A broad transmission spectrum at room temperature is shown in Fig.\ \ref{Fig_Transmission_exp_and_sim}.
%%%%%%%%%%%%%%%%%%%%%%%%%%%%%%%%%%%%%%%%%%%%%%%%%%%%%%%%%%%%%%%%%%%%%%%%%%%%
%%%%%%%%%%%%%%%%%%%%%%%%%%%%%%%%%%%%%%%%%%%%%%%%%%%%%%%%%%%%%%%%%%%%%%%%%%%%
Three regions of increased transmission reveal the eigenfrequencies of the cavity.
By comparing them with the simulations from Sec.\ \ref{Sec_FEM_simulations_resonance_frequency}, we clearly identify the isolated resonance around $\SI{34.383}{\mega\hertz}$ as the QM.
The measured eigenfrequency deviates by $\approx \SI{468}{\kilo\hertz}$ from the simulation result (see Tab.\ \ref{Tab_SRF_cavity_summary_eigenfrequency_simulations}).
This discrepancy is much larger than the estimated uncertainties of experiment and simulation, which could be explained by the effect of RF couplers and the external
electronic circuit, both neglected in the simulation.
The deviations of the other peaks from the simulation are larger than for the QM as a result of the simplified geometry.
Narrow scans of the reflection and transmission response function of the QM are displayed in Fig.\ \ref{Fig_Transmission_and_reflection_measured}.
%%%%%%%%%%%%%%%%%%%%%%%%%%%%%%%%%%%%%%%%%%%%%%%%%%%%%%%%%%%%%%%%%%%%%%%%%%%%
%%%%%%%%%%%%%%%%%%%%%%%%%%%%%%%%%%%%%%%%%%%%%%%%%%%%%%%%%%%%%%%%%%%%%%%%%%%%
For determining the free, i.e.\ unperturbed parameters of the cavity, the input power was lowered to $\SI{-80}{dBm}$ and $\SI{-70}{dBm}$ for the transmission and reflection scans, respectively, and the data fitted with the model from Eq.\ \ref{eq_s_param_general}.

In general, only the reflection measurement yields $Q_0$, while the transmission spectrum delivers $Q$.
From the reflection data, we obtain a resonance frequency of $\SI{34.5221608\pm 0.0000006}{\mega\hertz}$ and $Q_0=\SI{2.43\pm0.02e5}{}$.
The coupling constant of the inductive coupler, defined by $k_i=Q_0/Q_i$, is $k_1 = \SI{2.820\pm0.009}{}$, which corresponds to overcritical coupling.

In transmission, a shifted resonance frequency of $\SI{34.522204\pm .000002}{\mega\hertz}$ and $Q=\SI{5.79\pm0.05e4}{}$ is measured.
Compared with the result obtained in reflection, as the capacitive probe is only weakly coupled to the cavity, $k_2 \ll 1$, losses by the inductive coupler dominate.
In this approximation, the unloaded quality factor becomes $Q_0\simeq Q(1+k_1) = \SI{2.21\pm 0.06 e5}{}$, in reasonable agreement with the reflection analysis considering the simplified description.

The cavity can be impedance-matched to an external $\SI{50}{\ohm}$ signal generator, i.e.\ $k=1$, to drive it efficiently.
In this case, the reflected power vanishes at resonance, minimizing the input power needed for a desired intra-cavity power.
Tuning the coupling strength is carried out \cite{grieser_entwicklung_1986} by adjusting the angle $\gamma$ between the inductive coupler and the magnetic field lines of the QM inside the cavity.
Hereby, the transformed resistance of the \textit{LCR} resonant circuit,
\begin{eqnarray}
R_\text{in} = \frac{1}{2} \omega_0 N^2 A_\text{c}^2 \frac{B_0^2}{W} Q_0 \cos^2\gamma,
\end{eqnarray}
depends on the number of windings $N$, the area $A_\text{c}$ of the coupler, the enclosed magnetic flux $B_0$, and the total energy stored in the cavity $W$.
We tested this method with a normal-conducting prototype cavity\cite{stark_phd_2020} and plan to apply it to the SCC.

%%%%%%%%%%%%%%%%%%%%%%%%%%%%%%%%%%%%%%%%%%%%%%%%%%%%%%%%%%%%%%%%%%%%%%%%%%%%
%%%%%%%%%%%%%%%%%%%%%%%%%%%%%%%%%%%%%%%%%%%%%%%%%%%%%%%%%%%%%%%%%%%%%%%%%%%%
\section{Operation as quadrupole-mass filter}
\begin{figure}[t]
\centering
\includegraphics[width=\columnwidth]{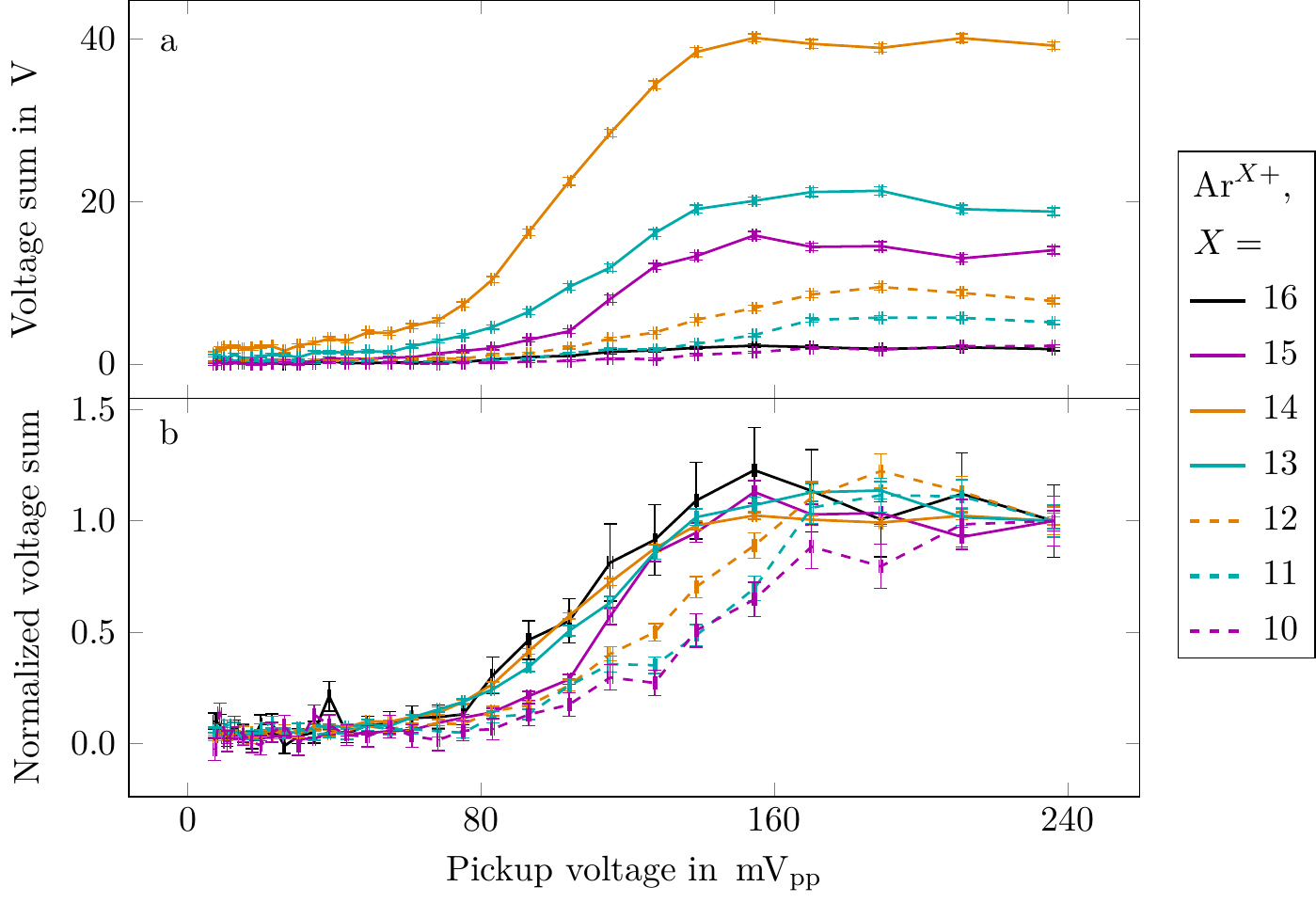}
\caption{Transmission efficiency for different Ar HCI as a function of the stored RF power quantified using the SCC pick-up voltage.
Points represent the background-subtracted TOF peak voltage sum averaged over 192 extraction cycles with their statistical uncertainties.
Reprinted from Ref.\ \onlinecite{stark_phd_2020}.}
\label{Fig_HCI_Transmission_Integrals}
\end{figure}

Since the SCC is designed to capture and store HCI from an external ion source, we first characterize the injection efficiency and HCI transmission by operating it as a quadrupole-mass filter radially confining the ion motion as a function of the intra-cavity RF power.
%%%%%%%%%%%%%%%%%%%%%%%%%%%%%%%%%%%%%%%%%%%%%%%%%%%%%%%%%%%%%%%%%%%%%%%%%%%%
\begin{figure}[t]
\centering
\includegraphics[width=\columnwidth]{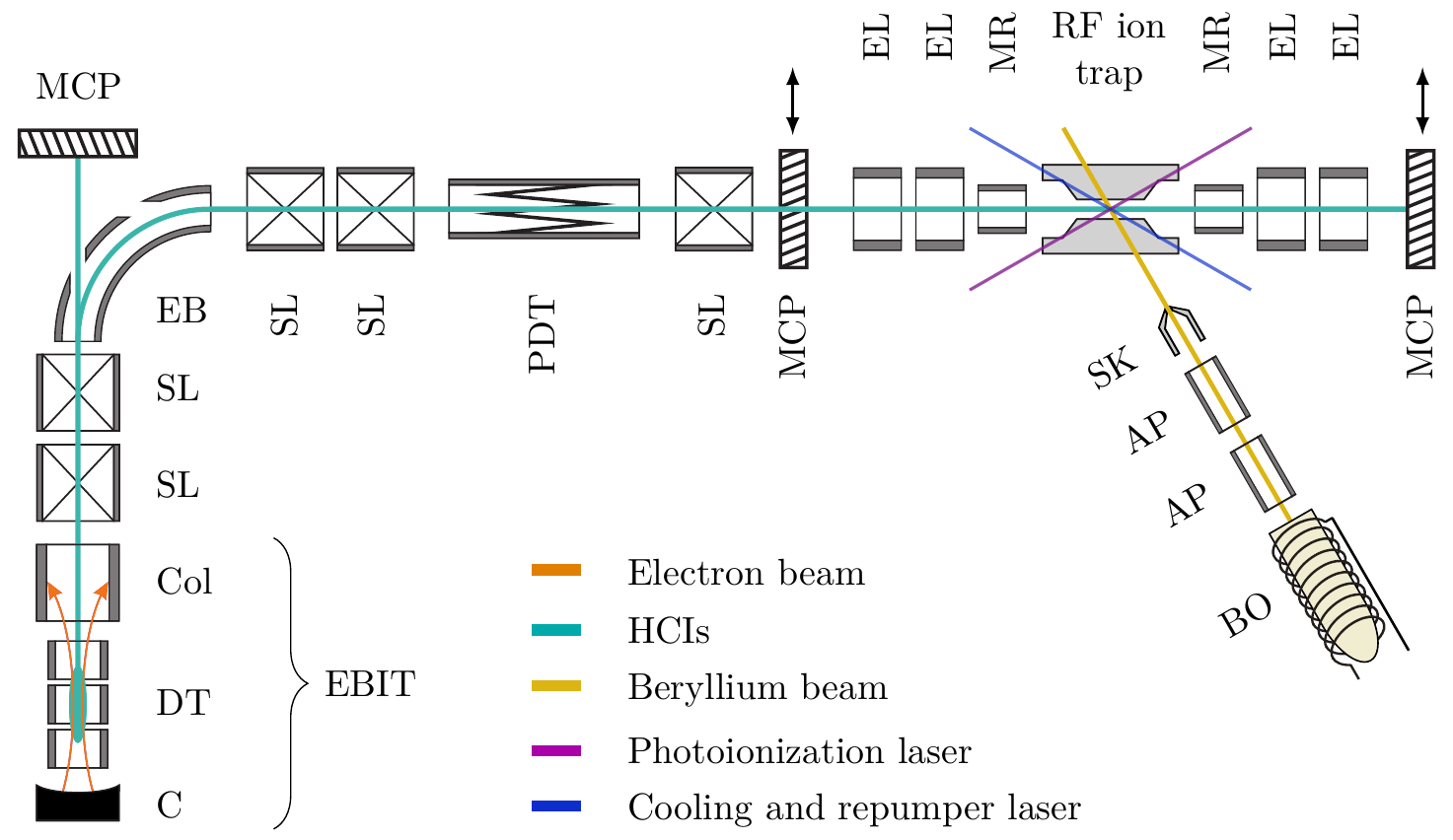}
\caption{Scheme of EBIT and HCI transfer beamline connected to the SCC.
C: Cathode, DT: Drift tubes, Col: Collector, SL: Sikler lens, EB: Electrostatic bender, PDT: Pulsed drift tubes, EL: Electrostatic lens, MR: Mirror electrode, SK: Skimmer, AP: Aperture, BO: Beryllium oven.
For details see text.
Adapted from Ref.\ \onlinecite{stark_phd_2020}.}
\label{Fig_EXP_Setup_Beamline_EBIT_PT}
\end{figure}
%%%%%%%%%%%%%%%%%%%%%%%%%%%%%%%%%%%%%%%%%%%%%%%%%%%%%%%%%%%%%%%%%%%%%%%%%%%%
%%%%%%%%%%%%%%%%%%%%%%%%%%%%%%%%%%%%%%%%%%%%%%%%%%%%%%%%%%%%%%%%%%%%%%%%%%%%
\begin{figure}[b]
\centering
\includegraphics[width=\columnwidth]{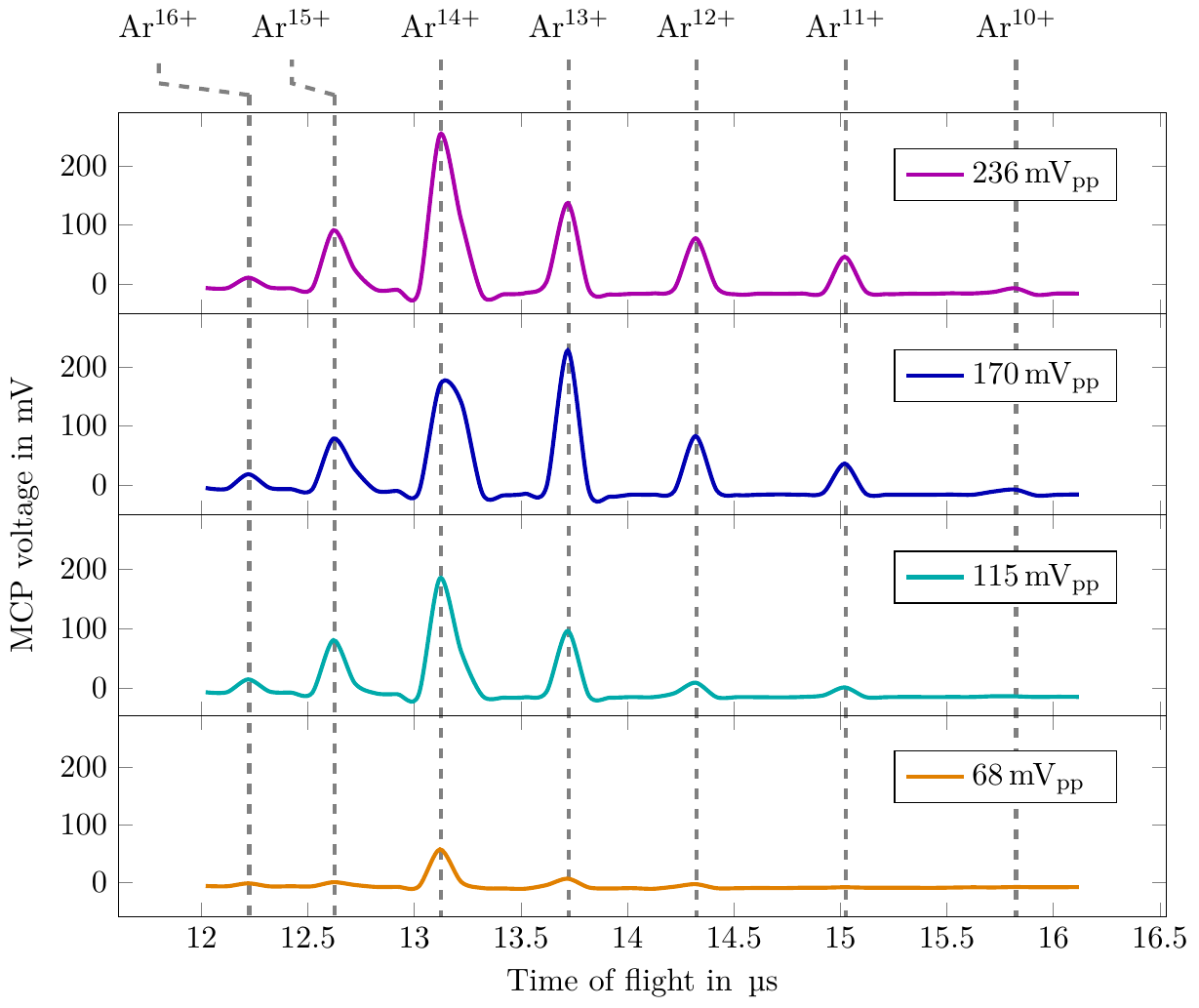}
\caption{TOF spectra of Ar HCI measured with an MCP detector behind the SCC as a function of stored RF power quantified using the SCC pick-up voltage.
Each spectrum represents the average of $192$ extraction cycles.
Reprinted from Ref.\ \onlinecite{stark_phd_2020}.}
\label{Fig_HCI_Transmission_3D}
\end{figure}
%%%%%%%%%%%%%%%%%%%%%%%%%%%%%%%%%%%%%%%%%%%%%%%%%%%%%%%%%%%%%%%%%%%%%%%%%%%%
We employ an EBIT as an HCI source connected to the SCC through a transfer beamline (Fig.\ \ref{Fig_EXP_Setup_Beamline_EBIT_PT}).
A Heidelberg compact EBIT,\cite{micke_heidelberg_2018} operated at $\SI{1.145}{\kilo\electronvolt}$ electron-beam energy produces argon ions in charge states up to $q/\text{e}=16$.
After a charge-breeding time of about $\SI{100}{\milli\second}$, the HCI are ejected in bunches with a kinetic energy of about $\SI{695}{\volt}\times q$.
During transfer, the different charge-to-mass $q/m$ species in the bunch separate according to their different time-of-flight (TOF), and are detected after passing the SCC with a microchannel plate (MCP) detector.
The beamline is operated with static potentials optimizing HCI transfer.
Under the given conditions, the fastest ions spend around $\SI{1}{\micro\second}$ inside the SCC, or approximately $35$ cycles of the RF field.
Typical TOF spectra in Fig.\ \ref{Fig_HCI_Transmission_3D} show peaks from $\text{Ar}^{10+}$ to $\text{Ar}^{16+}$ ions.
Their relative amplitudes cannot be directly compared, since these depend on their specific EBIT yields, the beamline transmission for a given $q/m$, and the sensitivity of the MCP to different charge states and ion impact energies.
The transmission efficiency for each individual charge state depends strongly on the SCC RF power, measured with the pick-up coupler.
At high power, the strong radial confinement of the ion motion improves the transmission, displayed in Fig.\ \ref{Fig_HCI_Transmission_Integrals} as the integral of each $q/m$ peak depending on the pickup voltage.
%%%%%%%%%%%%%%%%%%%%%%%%%%%%%%%%%%%%%%%%%%%%%%%%%%%%%%%%%%%%%%%%%%%%%%%%%%%%
%%%%%%%%%%%%%%%%%%%%%%%%%%%%%%%%%%%%%%%%%%%%%%%%%%%%%%%%%%%%%%%%%%%%%%%%%%%%
The efficiency increases with RF power until it saturates for most charge states above $\SI{180}{\milli\volt}_\text{pp}$ pickup voltage.
As expected for stable radial ion motion inside the SCC (see Eq.\ \ref{Eq_Stability_Condition_Paul_trap}), higher charge states show better transmission already at small RF power.
This measurement proves the stable radial confinement of HCI within the SCC.
The next step will be their deceleration and retrapping by Coulomb interaction with trapped laser-cooled ${}^9\textrm{Be}^+$ ions.

%%%%%%%%%%%%%%%%%%%%%%%%%%%%%%%%%%%%%%%%%%%%%%%%%%%%%%%%%%%%%%%%%%%%%%%%%%%%
%%%%%%%%%%%%%%%%%%%%%%%%%%%%%%%%%%%%%%%%%%%%%%%%%%%%%%%%%%%%%%%%%%%%%%%%%%%%
\section{Trapping of ${}^9\text{Be}^+$ ions}

Our first trapped-ion experiments with the setup sketched in Fig.\ \ref{Fig_EXP_Setup_Beamline_EBIT_PT} used ${}^9\text{Be}^+$ ions produced within the trap region by photoionization of ${}^9\text{Be}$ atoms from an atomic beam.
Because of their kinetic energies of $\approx \SI{140}{\milli\electronvolt}$, they are instantly captured by the SCC.
Subsequent Doppler cooling brings their temperature down to the $\SI{}{\milli\kelvin}$ range.
The effusive thermal Be beam emanates from an oven\cite{schmoger_kalte_2017} heated to $T=\SI{1250}{\kelvin}$ that is located at $\SI{0.93}{\metre}$ from the SCC and separated from it by two differentially pumped vacuum stages.
The Be beam is collimated to a diameter of $\SI{800}{\micro\metre}$ at the trap center for avoiding surface contamination of the superconducting electrodes.
There, it crosses the photoionization laser\cite{lo_all-solid-state_2014} at $90^{\circ}$, reducing the first-order Doppler shift to the MHz range.
Two-photon resonance-enhanced ionization proceeds through the $2\text{s}^1\text{S}_0 \rightarrow 2\text{p}^1\text{P}_1$ transition at $\SI{235}{\nano\metre}$.
With a $1/e^2$ laser-beam diameter of $\SI{250}{\micro\metre}$ at the SCC center, we could load the trap at laser powers above $\SI{80}{\micro\watt}$.
For Doppler cooling of ${}^9\text{Be}^+$ we use the strong ${}^2\text{S}_{1/2} (F=2)\rightarrow {}^2\text{P}_{3/2}$ transition at $\SI{313}{\nano\metre}$.
The required laser\cite{wilson_750-mw_2011} beam enters horizontally at an angle of $30^{\circ}$ to the trap axis, thus cooling both, axial and radial modes.
Perfect circular polarization would result in a closed cooling cycle when using a co-linear bias magnetic field, but in the present experiments we used the Earth magnetic field to define the quantization axis.
Therefore, some population is optically pumped into the $F=1$ hyperfine sublevel of the ground state, requiring a separate repumper beam detuned by $\SI{1.25}{\giga\hertz}$ from the cooling transition, with the same polarization and propagation axis as the cooling laser.
%%%%%%%%%%%%%%%%%%%%%%%%%%%%%%%%%%%%%%%%%%%%%%%%%%%%%%%%%%%%%%%%%%%%%%%%%%%%
\begin{figure}[t]
\centering
\includegraphics[width=\columnwidth]{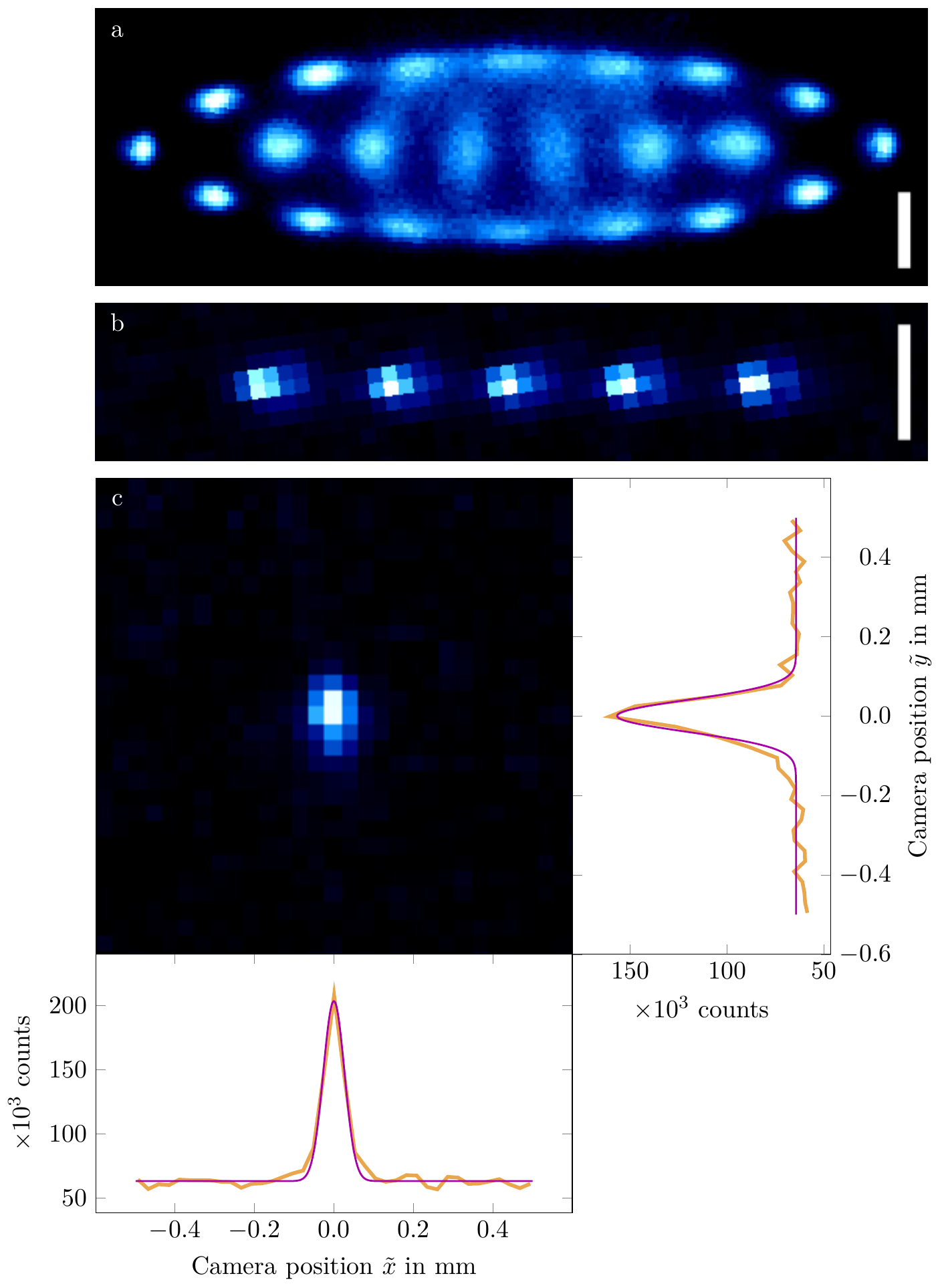}
\caption{Images of Doppler-cooled ${}^9\text{Be}^+$ ion ensembles confined inside the SCC.
(\textbf{a}, \textbf{b}) White bar: Scale of \SI{200}{\micro\metre} at the focal plane of the camera.
(\textbf{c}) Gaussian fits (magenta) to the projected intensity distribution of a single ion (orange) with standard deviations $\sigma_{\tilde{x}} = \SI{26.6\pm0.7}{\micro\metre}$ and $\sigma_{\tilde{y}}=\SI{40\pm2}{\micro\metre}$ at the camera.
Note the $2\times 2$ binning in (\textbf{b}) and (\textbf{c}).
}
\label{Fig_Ion_crystal}
\end{figure}
%%%%%%%%%%%%%%%%%%%%%%%%%%%%%%%%%%%%%%%%%%%%%%%%%%%%%%%%%%%%%%%%%%%%%%%%%%%%

Fig.\ \ref{Fig_Ion_crystal} shows images of Doppler-cooled ${}^9\text{Be}^+$ Coulomb crystals, magnified $M \approx 10$ times as defined by the chosen position of the $\SI{40}{\kelvin}$ asphere.
In these proof-of-principle experiments, the trap is typically operated with an RF input level of $\SI{21}{dBm}$ corresponding to RF amplitudes of $V_\text{RF}\approx\SI{105.6}{\volt}$ at the quadrupole electrodes, producing ion ensembles such as those shown in Fig.\ \ref{Fig_Ion_crystal}(b).
With axial DC potentials around $\SI{1}{\volt}$, the ${}^9\text{Be}^+$ ions are stored at secular frequencies of $\omega_z/2\pi \simeq \SI{209}{\kilo\hertz}$ and $\omega_r/2\pi \simeq \SI{409}{\kilo\hertz}$, calibrated by motional excitation of a single ${}^9\text{Be}^+$ ion confined inside the SCC.
Due to RF power dissipation of some elements, the trap heats up by $\sim\SI{1}{\kelvin}$ at these input levels, limiting the radial secular frequencies to about $\SI{500}{\kilo\hertz}$.
Eliminating these RF loss mechanisms will allow for motional frequencies in the MHz range as required for the application of QLS. We are currently identifying wiring parts that experience ohmic heating at higher RF input powers to further reduce those losses.

%%%%%%%%%%%%%%%%%%%%%%%%%%%%%%%%%%%%%%%%%%%%%%%%%%%%%%%%%%%%%%%%%%%%%%%%%%%%
%%%%%%%%%%%%%%%%%%%%%%%%%%%%%%%%%%%%%%%%%%%%%%%%%%%%%%%%%%%%%%%%%%%%%%%%%%%%
\section{Conclusion}\label{sec:VI}

We have introduced and commissioned a novel cryogenic ion trap employing a superconducting cavity which confines ions within the RF field of its electric quadrupole mode.
Its quality factor around $\SI{2.3e5}{}$ is two to three orders of magnitude higher than values reported for normal\cite{siverns_application_2012} and superconducting\cite{Poitzsch_1996} step-up resonators connected to cryogenic Paul traps, and may be further increased in the near future, as losses due to trapped magnetic flux\cite{vallet_flux_1992, gittleman_pinning_2003} inside the cavity walls or locally enhanced RF dissipation are eliminated.
Proof-of-principle operation showed a large acceptance for injected HCI and stable confinement of laser-cooled ${}^9\text{Be}^+$ Coulomb crystals.
The cavity bandpass strongly suppresses white noise from the RF power supply at the trap electrodes.
Spectral components at the secular frequencies $\omega_i$ or their sidebands around the trap drive at $\Omega \pm \omega_i$, both of which can cause motional heating of the ion, are reduced by a factor of $>10^{4}$.
This should result in extremely small motional heating rates below values reported for other cryogenic Paul trap experiments.\cite{brownnutt_ion-trap_2015,johnson_active_2016, brandl_cryogenic_2016,leopold_cryogenic_2019, dubielzig2020ultralow}
Such small rates around a few quanta per second are measured using sideband thermometry,\cite{turchette_heating_2000} requiring preparation of the ions in their motional ground state.
We will in the near future implement the scheme described in Ref.\ \onlinecite{leopold_cryogenic_2019} with a laser system currently being developed.

Using the recently established techniques for trapping,\cite{schmoger_deceleration_2015} cooling,\cite{schmoger_deceleration_2015, micke_coherent_2020} and subsequent interrogation of HCI by QLS,\cite{micke_coherent_2020} this new apparatus, in combination with the XUV frequency comb,\cite{nauta_towards_2017, Nauta_VMI_2020, nauta2020xuv} promises the implementation of XUV frequency metrology based on HCI.

%Using the newly developed techniques for trapping,\cite{schmoger_deceleration_2015} cooling,\cite{schmoger_deceleration_2015, micke_coherent_2020} and subsequent interrogation of HCI using QLS,\cite{micke_coherent_2020} we aim for a precision of state-of-the-art single-ion optical frequency standards,\cite{ludlow_optical_2015, huntemann_single-ion_2016, brewer_quantum-logic_2019} while possibly improving upon the reported accuracies.
%In combination with the VUV frequency comb\cite{nauta_towards_2017, Nauta_VMI_2020} this unique apparatus might ultimately lead the way towards VUV frequency metrology based on HCI.

\section*{Data availability statement}
The data that support the findings of this study are available from the corresponding author upon reasonable request.

%%%%%%%%%%%%%%%%%%%%%%%%%%%%%%%%%%%%%%%%%%%%%%%%%%%%%%%%%%%%%%%%%%%%%%%%%%%%
%%%%%%%%%%%%%%%%%%%%%%%%%%%%%%%%%%%%%%%%%%%%%%%%%%%%%%%%%%%%%%%%%%%%%%%%%%%%
\begin{acknowledgments}
We acknowledge the MPIK engineering design office led by Frank Müller, the MPIK mechanical workshop led by Thorsten Spranz, and the MPIK mechanical apprenticeship workshop led by Stefan Flicker and Florian Säubert for their expertise and the fabrication of numerous parts as well as the development of sophisticated fabrication procedures of complex parts.
For their technical support, we also thank Thomas Busch, Lukas Dengel, Nils Falter, Christian Kaiser, Oliver Koschorreck, Steffen Vogel and Peter Werle.
We acknowledge J. Iversen, D. Reschke, and L. Steder for support and discussions.
This project receives funding from the the Max-Planck Society, the Max-Planck–Riken–PTB–Center for Time, Constants and Fundamental Symmetries, the European Metrology Programme for Innovation and Research (EMPIR), which is co-financed by the Participating States and from the European Union’s Horizon 2020 research and innovation programme (project number 17FUN07 CC4C), and the Deutsche Forschungsgemeinschaft (DFG, German Research Foundation) through the collaborative research centre SFB 1225 ISOQUANT, through Germany’s Excellence Strategy - EXC-2123 QuantumFrontiers - 390837967, and through SCHM2678/5-1.
\end{acknowledgments}

%%%%%%%%%%%%%%%%%%%%%%%%%%%%%%%%%%%%%%%%%%%%%%%%%%%%%%%%%%%%%%%%%%%%%%%%%%%%
%%%%%%%%%%%%%%%%%%%%%%%%%%%%%%%%%%%%%%%%%%%%%%%%%%%%%%%%%%%%%%%%%%%%%%%%%%%%
\bibliography{SRFtrap_for_HCI}

\end{document}